\documentclass[10pt,epsf,preprint]{aastex}

\input psfig

\catcode`\@=11
\def\gsim{\ifmmode{\mathrel{\mathpalette\@versim>}}
    \else{$\mathrel{\mathpalette\@versim>}$}\fi}
\def\lsim{\ifmmode{\mathrel{\mathpalette\@versim<}}
    \else{$\mathrel{\mathpalette\@versim<}$}\fi}
\def\@versim#1#2{\lowerpaper2_9mar01.tex 2.9truept \vbox{\baselineskip 0pt \lineskip 
    0.5truept \ialign{$\m@th#1\hfil##\hfil$\crcr#2\crcr\sim\crcr}}}
\catcode`\@=12

\def\listitem{\par \hangindent=50pt\hangafter=1
     $\ $\hbox to 20pt{\hfil $\bullet$ \hfil}}

\def\puncspace{\ifmmode\,\else{\ifcat.\C{\if.\C\else\if,\C\else\if?\C\else%
\if:\C\else\if;\C\else\if-\C\else\if)\C\else\if/\C\else\if]\C\else\if'\C%
\else\space\fi\fi\fi\fi\fi\fi\fi\fi\fi\fi}%
\else\if\empty\C\else\if\space\C\else\space\fi\fi\fi}\fi}
\def\SP{\let\\=\empty\futurelet\C\puncspace}

\def\kms{kms$^{-1}$}
\def\h-1{$h^{-1}$}

\def\void#1{{}}

\def\h1{$h^{-1}$}

\def\kms{kms$^{-1}$ }
\def\etal{et al.\SP}
\def\eg{e.g., \,}

\def\lsim{~\rlap{$<$}{\lower 1.0ex\hbox{$\sim$}}}
\def\gsim{~\rlap{$>$}{\lower 1.0ex\hbox{$\sim$}}}
\def\dnsig{$D_n-\sigma$\SP}
\def\mg2sig{Mg$_2-\sigma$\SP}

\begin{document}

\title{Redshift-distance Survey of Early-type Galaxies: The
$D_n-\sigma$ Relation 
\footnote{Based on observations at Complejo Astronomico El
Leoncito (CASLEO), operated under agreement between the Consejo
Nacional de Investigaciones Cient\'\i ficas de la Rep\'ublica
Argentina and the National Universities of La Plata, C\'ordoba and San
Juan; Cerro Tololo Interamerican Observatory (CTIO), operated by the
National Optical Astronomical Observatories, under AURA; European
Southern Observatory (ESO), partially under the ESO-ON agreement; Fred
Lawrence Whipple Observatory (FLWO); Observat\'orio do Pico dos Dias,
operated by the Laborat\'orio Nacional de Astrof\'\i sica (LNA); and
the MDM Observatory on Kitt Peak.} }

\author{M. Bernardi\altaffilmark{1},
M. V. Alonso\altaffilmark{2}, L. N. da Costa\altaffilmark{3,}\altaffilmark{4}, 
C. N. A. Willmer\altaffilmark{4,}\altaffilmark{5}}

\author{G. Wegner\altaffilmark{6},
P. S. Pellegrini\altaffilmark{4}, C. Rit\'e\altaffilmark{4},
M. A. G. Maia\altaffilmark{4}}

\affil{\altaffiltext{1}{The University of Chicago, 5640 South Ellis
Avenue, Chicago, IL 60637, USA}}

\affil{\altaffiltext{2}{Observatorio Astr\'onomico de
C\'ordoba,  Laprida  854, C\'ordoba, 5000, Argentina}}

\affil{\altaffiltext{3}{European Southern Observatory,
Karl-Schwarzschild Strasse 2, D-85748 Garching,
Germany}}

\affil{ \altaffiltext{4} {Departamento de Astronomia,
Observat\'orio Nacional, Rua Gen. Jos\'e Cristino
77, Rio de Janeiro, R.J., 20921, Brazil}}

\affil{ \altaffiltext{5}{UCO/Lick Observatory, University of California,
1156 High Street, Santa Cruz,  CA 95064, USA}}

\affil{\altaffiltext{6}{Department of Physics \& Astronomy, Dartmouth
College, Hanover, NH  03755-3528, USA}}


\begin{abstract}

In this paper $R-$band photometric 
and velocity dispersion measurements for a
sample of 452 elliptical and S0 galaxies in 28 clusters are used to
construct a template $D_n-\sigma$ relation. This template relation is
constructed by combining the data from the 28 clusters, under the
assumption that galaxies in different clusters have similar 
properties.  The photometric and spectroscopic data used consist of 
new as well as published measurements, converted to a common system,
as presented in a accompanying paper.  The
resulting direct relation, corrected for incompleteness bias, is
$\log{D_n} =1.203 \log{\sigma} + 1.406$;
the zero-point has been defined by requiring distant clusters to be at
rest relative to the CMB. This zero-point is consistent with the value
obtained by using the distance to Virgo as determined by the Cepheid
period-luminosity relation. This new $D_n-\sigma$ relation leads to a
peculiar velocity of $-72 \pm 189$ kms$^{-1}$ for the Coma
cluster.  The scatter in the distance relation
corresponds to a distance error of about 20\%, comparable to the
values obtained for the Fundamental Plane relation.  Correlations
between the scatter and residuals of the $D_n-\sigma$ relation with
other parameters that characterize the cluster and/or the galaxy
stellar population are also analyzed.  The direct and inverse 
relations presented here have been used in recent studies of the 
peculiar velocity field mapped by the ENEAR all-sky sample.

\end {abstract}
\clearpage

\section{Introduction}
\label{intro}

Present-day elliptical galaxies form a remarkably homogeneous class of
objects which obey scaling relations involving their structural and
dynamical properties. Indeed, elliptical galaxies are known to populate
the so-called fundamental plane (FP, Djorgovski \& Davis 1987; Dressler
\etal 1987), in a three-dimensional space defined by the surface
brightness $\mu_e$, the effective radius $r_e$, and internal velocity
dispersion $\sigma$.  Therefore, by choosing an appropriate combination
of parameters, a tight relation between distance--dependent and
independent quantities can be found; the $D_n-\sigma$ relation is such a
relation (Dressler \etal 1987), 
where $D_n$ is the physical scale of the galaxy defined at a
specified surface brightness level ($D_n\equiv d_n \times R$, where
$d_n$ is a measure of the angular size of the galaxy and $R$ is the
distance of the galaxy). The existence of such scaling relations
provides an important tool for studying the properties of the stellar
populations and the evolution of ellipticals 
(\eg J{\o}rgensen, Franx, \& Kj{\ae}rgaard 1996;
Franx \etal 1997), and for constraining models of spheroidal formation
(\eg Bressan, Chiosi, \& Fagotto 1994; Baugh, Cole, \& Frenk 1996). 
Furthermore, these
relations provide the means of measuring relative distances to
early-type galaxies.  This is the primary goal of this work. 
At the present time, some doubts remain whether these relations 
depend on the environment (\eg Gibbons, Fruchter, \& Bothun 2001); if they do, 
this would lead to the measurement of spurious motions.  Also, 
until recently, it was unclear how distances derived using $D_n-\sigma$  
related to those measured by the Tully-Fisher relation 
(Scodeggio 1997; Scodeggio, Giovanelli, \& Haynes 1997).  

The FP and the $D_n-\sigma$ scaling relations are not entirely
equivalent (J{\o}rgensen, Franx, \& Kj{\ae}rgaard 1993) 
and the $D_n-\sigma$ relation is
expected to be less accurate if the range of galaxy sizes is large
(Kelson \etal 2000). In addition, it has been claimed that the scatter
around the FP is smaller, suggesting that $D_n-\sigma$ distances are
less accurate (J{\o}rgensen \etal 1993). However, recent studies do
not seem to support these claims (J{\o}rgensen \etal 1996; D'Onofrio
\etal 1997; Hudson \etal 1997). While the FP relation is usually used
for detailed cluster studies, for large samples such as the
magnitude-limited, all-sky sample of early-type galaxies (ENEAR, da
Costa \etal 2000a) presented below, the number of galaxies with
available $d_n$ measurements is $\sim 50$\% larger than the number
with available FP measurements. Therefore, since the primary goal of
this project has been to estimate galaxy distances and derive the
peculiar velocity field, we have focused our attention on the derivation
of a template $D_n-\sigma$ relation. A similar analysis for the FP
relation will be presented in a future paper.  

This work uses the ENEARc sample of early-type galaxies in 28 clusters
presented by Bernardi \etal (2001; hereafter B01).  This sample combines
data available in the literature with new measurements converted into a
common system thanks to the effort of securing new measurements for a
large number of galaxies in common with previously available samples. 
Another important feature of the sample is that cluster membership  was
carried out using groups identified in complete redshift surveys of the
nearby universe, thereby leading to a more systematic assignment than
was possible in earlier work.

In this paper we obtain \dnsig fits for each of the 28 clusters 
accounting for various possible biases.  
We then combine the sample to construct a global template relation
under the assumption that early-type galaxies in different clusters 
are similar.  We also study the residuals with respect to the template 
to investigate, a posteriori, the accuracy of this assumption. 
The resulting relation  is used to compute peculiar velocities 
of clusters as well as of the ENEAR all-sky sample of early-type 
galaxies.  Both these samples have been used to measure the 
bulk flow velocity, to set constraints on cosmological parameters 
and to characterize the velocity field and mass distribution in
the local universe (da Costa \etal 2000b; Borgani \etal 2000b; Nusser
\etal 2001; Zaroubi \etal 2001).

The paper is organized as follows: in Section~\ref{sample} the data
set with the individual galaxy parameters is described. In
Section~\ref{relation} we present the calibration of the direct
(forward) \dnsig relation. We quantify the selection bias which, if
not corrected for, can lead to an erroneous determination of the
distance relation coefficients and its scatter. This is critical for
studies of the cosmic flow field and the properties of early-type
galaxies. The direct relation has been applied to compute distances
for the galaxy sample used in da Costa \etal (2000b) and Zaroubi \etal
(2001).  Also shown are the parameters for the inverse relation
obtained by regressing on the distance-independent quantity log
($\sigma$); this inverse relation has been used in the analysis of the
peculiar velocity field of clusters and ``field'' galaxies in redshift
space (Borgani \etal 2000b; Nusser \etal 2001).  The measured
distances and peculiar velocities for the ENEARc sample are reported
in Section~\ref{vpec}.  In Section~\ref{implications} we look 
for potential systematic effects which may invalidate the underlying
assumption that galaxies in different clusters are similar. 
Finally, in Section~\ref{summary} we present a brief summary of 
our results.

\section{The Cluster Sample}
\label{sample}

The spectroscopic and photometric parameters for the 452 galaxies in 
28 clusters used here are presented in B01, where we describe the
selection of the cluster sample and membership assignments. The clusters
we consider in the present study span the redshift range $1000
\lsim cz \lsim 11,000$ kms$^{-1}$, covering both equatorial hemispheres.
The characteristic parameters (mean redshift, size and velocity
dispersion) of nearly all clusters were computed from the analysis
of ``groups'' identified using objective friend-of-friends algorithms
applied to complete redshift surveys (see B01). Exceptions include the
Centaurus complex, three clusters previously studied by J{\o}rgensen,
Franx, \& Kj{\ae}rgaard (1995a, 1995b) and J{\o}rgensen (1997)
(A539, AS639, and A3381), and two observed by
Smith \etal (1997) (7S21 and A347). The parameters adopted for these
cases and the reasons for including them are discussed in B01. Using the
identified groups as signposts for clusters, galaxies fainter than the
ENEAR magnitude limit ($m_B \sim14.5$) were considered members by adopting
well-defined position and kinematic criteria which should minimize
errors in the membership assignment. About 2\% of the galaxies
previously assigned to clusters were found not to be members according
to the membership criteria adopted.

The data set of the ENEARc sample is a compilation including new
photometric and spectroscopic measurements obtained as part of this
program as well as data previously reported in the literature.  In B01
we presented the photometric and spectroscopic measurements for 640
individual cluster galaxies, including new measures of the photometric
parameter $d_n$ for 348 galaxies, new spectroscopic measurements of
redshift, velocity dispersion and the Mg$_2$ index for 229
galaxies. Our new data for cluster galaxies have been combined with
those in the literature by converting all measurements to a common
system (see B01). This was possible by securing observations for a
representative number of galaxies in common with other samples, thus
allowing the definition of conversion relations. Data from the
literature come from Dressler (1987), Lucey \&
Carter (1988), Faber \etal (1989), Dressler, Faber, \& Burstein
(1991), J{\o}rgensen
\etal (1995a, 1995b), Lucey \etal (1997), and Smith \etal (1997).  A
detailed description of the new $R-$band imaging data and parameters,
including the total magnitude, the effective radius $r_e$, mean
surface brightness within this radius $\mu_e$ and disk-to-bulge ratio
$D/B$ will be presented in Alonso \etal (2001) while the spectroscopic
data will be presented by Wegner \etal (2001).

In constructing the \dnsig relation, we exclude 188 galaxies 
(see B01, Table~8) either because they present photometric or spectroscopic
features typical of later type galaxies (\eg presence of arms, bar, dust
lane, emission lines) or because the measured parameters could be
affected due to contamination by nearby galaxies or stars. 
Pruning the sample in this way decreases the scatter ($\sim 5$\%) but
leaves both the slope and zero-point essentially unchanged.  

The selection and completeness of the sample of early-type galaxies 
in clusters is not very well defined because: a) it varies from cluster 
to cluster; and b) each cluster is a compilation of galaxies taken from 
different sources. The sample of cluster galaxies includes: 
i) all early-type galaxies brighter than 
$m_R < 14.5$ (since they were extracted from complete
magnitude-limited catalogs, see da Costa et al. 2000a); 
ii) fainter early-types with 
photometric and spectroscopic data available in the literature (see B01). 
Furthermore, for any given $S/N$ and resolution, there is a lower limit
below which the velocity dispersion measurements are unreliable.
Since a cluster may have measurements taken from different sources,
this lower limit is not well defined either. For example, for some sources
in the literature only measurements of $\sigma > 100$ kms$^{-1}$
are available; for our data this limit can be as low as $45$ kms$^{-1}$ 
due to the higher resolution used (see Wegner et al. 2002).
We have checked that the results presented below are not significantly 
dependent on the adopted velocity dispersion limit.

\section {Determining the distance relation}
\label{relation}

\subsection {The method}
\label{method}

A galaxy's angular size varies inversely as $R$, its co-moving 
distance. If $d_n$ is the measured size of the galaxy on the sky, 
then $1/d_n$ is a measure of its distance.  
The central velocity dispersion of a galaxy $\sigma$ is 
expected to be correlated with its physical size 
$D_n\equiv d_n\times R$ (e.g. Dressler et al. 1987).  
If we measure both $d_n$ and $\sigma$, then the basic distance 
indicator becomes
\begin{equation}
\log{R} = a\log{\sigma} - \log{d_n} + b ,
\label{eq:dnsig1}
\end{equation}
where $a$ represents the scaling of velocity dispersion with 
size:  $D_n\propto\sigma^a$.  
Here both $R$ and $\sigma$ are in units of kms$^{-1}$, and $d_n$ 
is expressed in units of 0.1~arcmin.  Define the quantities 
$y\equiv\log D_n + \log h$, where $h=H_o$/(100 kms$^{-1}$Mpc$^{-1}$), 
and $x\equiv \log \sigma$.  Then the distance relation is
\begin{equation}
y = ax + b .
\label{eq:dnsig}
\end{equation}

The slope of equation~(\ref{eq:dnsig}) is usually determined using
cluster galaxies, because they can be assumed to be equally distant 
and the uncertainty in the estimated distances falls as $1/{\sqrt N}$. 
If a distance relation which does not depend on cluster properties 
exists, it can be determined by combining the data from all available 
clusters to produce a standard template relation. 
Although the peculiar velocity field is unknown, combining many 
different clusters should improve the statistical accuracy of the 
slope and zero-point.  (Note that here we are only assuming that 
early-type galaxies in different clusters are similar; we are not 
addressing the possible differences between galaxies in clusters 
and in regions of lower density.)  

Here, the template parameters---zero-point, slope, and relative 
motions of each cluster---are determined simultaneously.  
Such a procedure has been adopted by a number of authors 
(Baggley 1996; Giovanelli \etal 1997, hereafter G97; Scodeggio 1997; 
Scodeggio \etal 1998; Colless \etal 2001) in determining the \dnsig, 
FP and TF relations.  Our notation below follows G97. 
The distance relation can be derived by either a direct (forward) 
or inverse linear regression fit, depending on whether the slope is
obtained using the distance-dependent $d_n$ or the 
distance-independent parameter $\sigma$ as the independent 
variable.  We study the direct first and the inverse later, in
Section~\ref{fits}.

The coefficients of the direct relation  
 $y = a_dx + b_d$ 
are determined as follows.  
We have $N_g$ galaxies in $N_{cl}$ clusters.
Let $(x_{ik}, y_{ik})$ denote the values of $x$ and $y$ for the 
$i$-th galaxy in the $k$-th cluster.  For the $k$-th cluster, 
the distance relation is  $y = a_kx + b_k$.  Our assumption that 
the distance relation does not depend on the properties of a 
cluster means that $a_k$ has the same value, $a_d$, for all 
clusters. If as a first guess one uses the observed radial velocity 
$cz$ as the ``distance'' 
($y\equiv\log D_n + \log h \equiv \log (d_n\times cz) + \log h$),
the zero-point $b_k$ is different for 
different clusters only because of their peculiar velocities 
relative to the Hubble flow:  
$b_k = b_d + \Delta_k$.  We would like to find those values of 
$a_d$, $b_d$ and $\Delta_k$ for which the scatter around the 
mean relation is minimized. Therefore, we minimize 
\begin{equation}
\label{eq:chi2} 
\chi^2 = \sum_{k=1}^{N_{cl}} \sum_{i=1}^{N_g(k)}
\Big\lbrack \frac {y_{ik}-(a_dx_{ik} + b_d + \Delta_k)} {\sigma_{ik}}
\Big\rbrack^2, 
\end{equation} 
with respect to the slope $a_d$, the zero-point $b_d$,  
and the relative offsets $\Delta_k$.  
Here $\sigma_{ik}$ is related to the measurement error in $D_n$ of 
the $i$-th galaxy in the $k$-th cluster.  

If ``distant'' clusters (which we define as being clusters beyond 
3000 kms$^{-1}$) are at rest relative to the CMB then the sum over 
their peculiar velocities should equal zero.  
Therefore, once $a_d$, $b_d$ and the $\Delta_k$'s have been found, 
we compute 
\begin{displaymath} 
 \sum_k N_g(k)\Delta_k/\sum_k N_g(k) 
\end{displaymath} 
where the sum is over the subset of ``distant'' clusters in our 
sample.  We then substract this value from each of the $\Delta_k$'s.  
In effect, this sets the overall zero-point of the distance relation.

As will be shown below, this condition turns out to be equivalent 
to requiring 
(i) the distance to Virgo equal that given by the Cepheid 
period-luminosity relation (Kelson \etal 2000), or
(ii) assuming that the Coma cluster is at rest.  

Formally, the equations above fully describe the procedure we 
use to determine the parameters which describe the distance relation.  
However, when working with real data, one must also consider possible 
sources of bias (for a review, see Strauss \& Willick 1995 and 
references therein), as described below.  

\subsection{Monte-Carlo Bias Correction}\label{monte}
For the direct relation, i.e., when fitting on the distance-dependent 
parameter, the most pernicious bias is that due to incompleteness. 
This bias leads to a shallower slope, a larger zero-point, and an 
underestimate of the scatter.  Although analytic bias-correction 
schemes have been proposed (Willick 1994), the assumptions made 
are hardly met by real data.  
This bias is particularly difficult to handle when the completeness 
varies from cluster to cluster, as is the case in our sample.  
Here we follow G97, Scodeggio (1997) and Scodeggio \etal (1998) 
and use a Monte-Carlo approach to estimate the bias correction, 
although this is not the only method that can be used 
(Wegner \etal 1996).  

As mentioned above, to estimate the bias, we must first know the
incompleteness in $D_n$ for each cluster.  This requires knowledge 
of the $D_n$ distribution function, the counterpart  of the luminosity
function.  Since this function is not directly available, two
approaches are possible:  ($i$) assume that a fair representation 
of this distribution is given by that of a nearby cluster which 
is complete; ($ii$) examine the correlation of $d_n$ with some other 
measure of the angular size of a galaxy, whose distribution is known.  
We adopt the second approach.  

Let $\theta_{25}$ denote the angular diameter enclosing an integrated 
surface brightness of 25~mag~arcsec$^{-2}$, and let $D_{25}$ 
denote the physical size obtained by multiplying this angular size 
by the distance to the galaxy.  The distribution of $D_{25}$ in the 
ESO-LV catalog (Lauberts \& Valentijn 1989) is 
\begin{equation} 
\phi (D_{25})\, dD_{25} \propto  
 \exp \left( \frac{-D_{25}}{D_{\star}}\right)\,\frac{dD_{25}}{D_{\star}} ,
\end{equation} 
with $D_{\star}=2610$ kms$^{-1}$ (Sodr\'e \& Lahav 1993).  
There is a tight correlation between $d_n$ and $\theta_{25}$
(e.g., Wegner et al. 1996), so the distribution of $D_n$ in a 
complete sample should be well approximated by 
\begin{equation}
\Phi(D_n) = \phi(D_{25})\,{dD_{25}\over dD_n}.
\label{eq:diamfunc}
\end{equation}
The results are insensitive to the exact shape of the diameter 
function (e.g., G97).  The ratio of the observed distribution 
of $D_n$ in a given cluster with the one expected for a complete 
sample provides an estimate of the completeness.  This ratio 
depends on $D_n$ differently for each cluster:  we call it the 
completeness $C_k(D_n)$.  Note that $C_k(D_n)$ varies between 
zero and one.  

Incompleteness leads to a bias in determining the distance 
indicator coefficients $a_d$ and $b_d$, which we estimate using 
the following Monte-Carlo approach.
In the first step, distances to clusters are approximated by using 
their redshifts, and the $\chi^2$ defined in equation~(\ref{eq:chi2}) 
is minimized.  This provides initial guesses for the slope, 
zero-point, peculiar velocities, and scatter $\epsilon$ around the 
mean relation.  The scatter $\epsilon$ may change with velocity 
dispersion, so we actually compute $\epsilon(x)$ in bins of $x$.  

For the $i$-th galaxy in the $k$-th cluster a bias correction 
$B_{ik}$ is obtained as follows.  A Gaussian zero mean unit 
variance random number $g$ is generated.  
This, with the coefficients $a_d$ and $b_d$ and the scatter 
$\epsilon(x_{ik})$, is used to compute 
$y^s_{ik} = a_d x_{ik} + b_d + g\epsilon(x_{ik})$.
This represents the value of $y$ the observed galaxy may have had.  
If this value of $y$ was too small, the galaxy would not have been 
observed.  The probability it would have been observed is proportional 
to the completeness $C_k(y^s_{ik})$.  Therefore, we generate a 
random number $u$ which is distributed uniformly between zero and one.  
The number $y^s_{ik}$ is accepted if $u\le C_k(y^s_{ik})$.  
We repeat this procedure until we have acccepted 500 values of 
$y^s_{ik}$ for each galaxy.  The mean of these values 
$\langle y^s_{ik}\rangle$ reflects the incompleteness of the 
real sample.  It thus allows a direct estimate of the bias:
\begin{equation}
B_{ik}= \langle y^s_{ik}\rangle - (a_d x_{ik} + b_d).
\label{eq:bias}
\end{equation}
This value is used to define corrected values 
\begin{equation}
y^c_{ik} = (a_d x_{ik} + b_d) - B_{ik}.  
\label{eq:bias2}
\end{equation}
These corrected values are inserted in equation~(\ref{eq:chi2});
minimizing yields new estimates of $a_d$, $b_d$, the $\Delta_{k}$s, 
and the scatter $\epsilon$.  The process is repeated until 
convergence is reached; applied to our data, this happens after 
about four iterations.  

\subsection{The Template Distance Relation: Fitting Parameters}
\label{fits}

\begin{figure} 
\centering 
\mbox{\psfig{figure=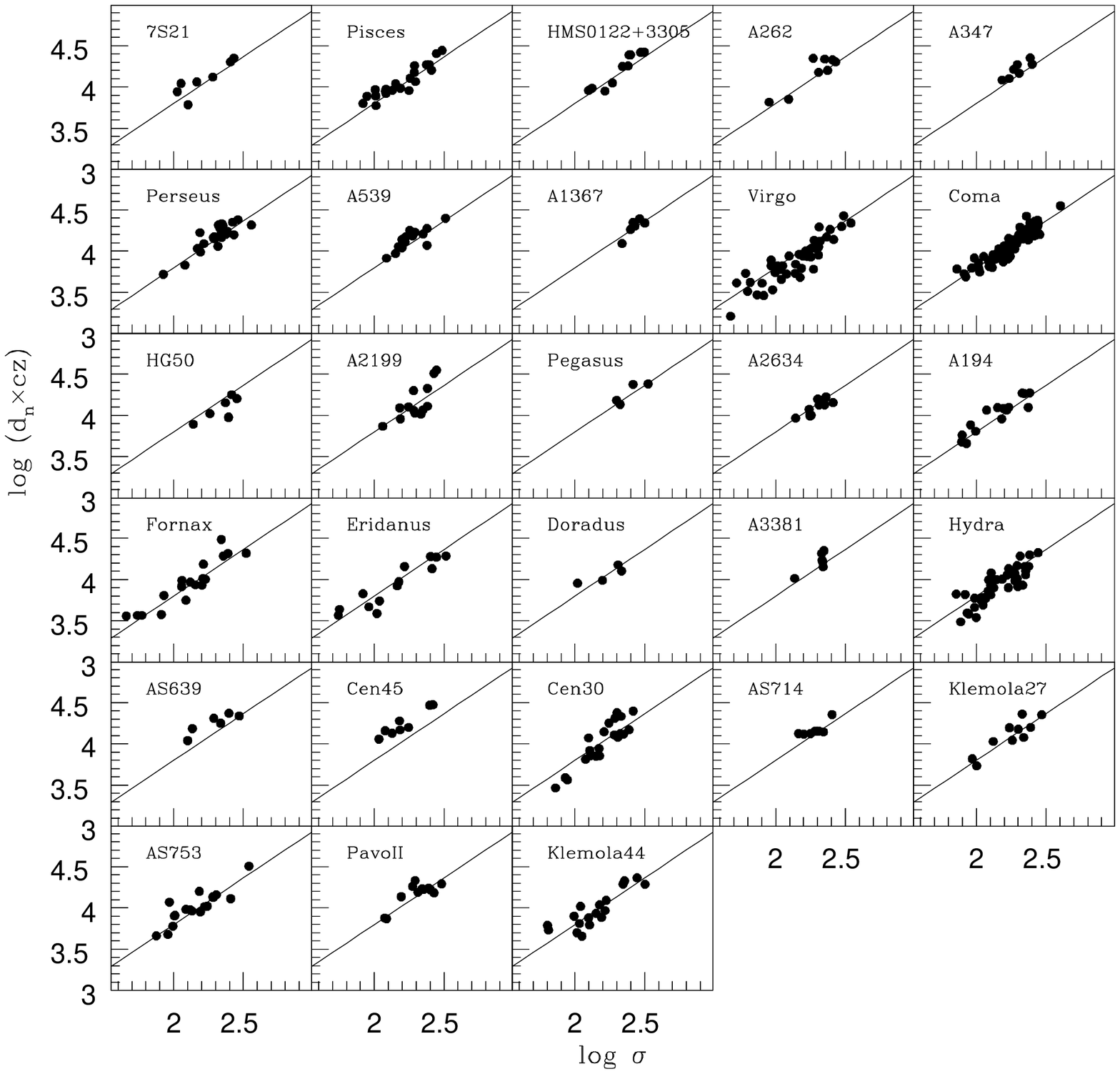,height=15truecm,bbllx=2truecm,bblly=5truecm,bburx=19truecm,bbury=23truecm }}
\caption{The product of $d_n$ (in 0.1~arcmin) and the redshift 
(in kms$^{-1}$) of each cluster galaxy is plotted versus its velocity 
dispersion. The solid line represents the best fit after minimizing
the $\chi^2$ defined in equation~(\ref{eq:chi2}) for the first time.}  
\label{fig:DS_groupcz}
\end{figure}

We apply the above procedure to the
cluster sample presented in Section~\ref{sample}. We start by assuming
that the clusters are at rest relative to the Hubble flow and that their
distances are given by the mean cluster redshift (see B01).
Figure~\ref{fig:DS_groupcz} shows the individual uncorrected cluster
data at the start of the iterative process. The solid line represents
the best fit after minimizing $\chi^2$ (equation~\ref{eq:chi2}) for the
first time.  Note that the number of galaxies in each cluster varies
dramatically and, for most groups, only the more luminous, high velocity
dispersion cluster galaxies are included in the  sample. Therefore, if
the selection bias correction is not applied, a significant bias exists
in the global template constructed using all clusters.
The relative offsets between those data points and the distance relation
reflect the relative motions of the clusters.

\begin{figure}
\centering
\mbox{\psfig{figure=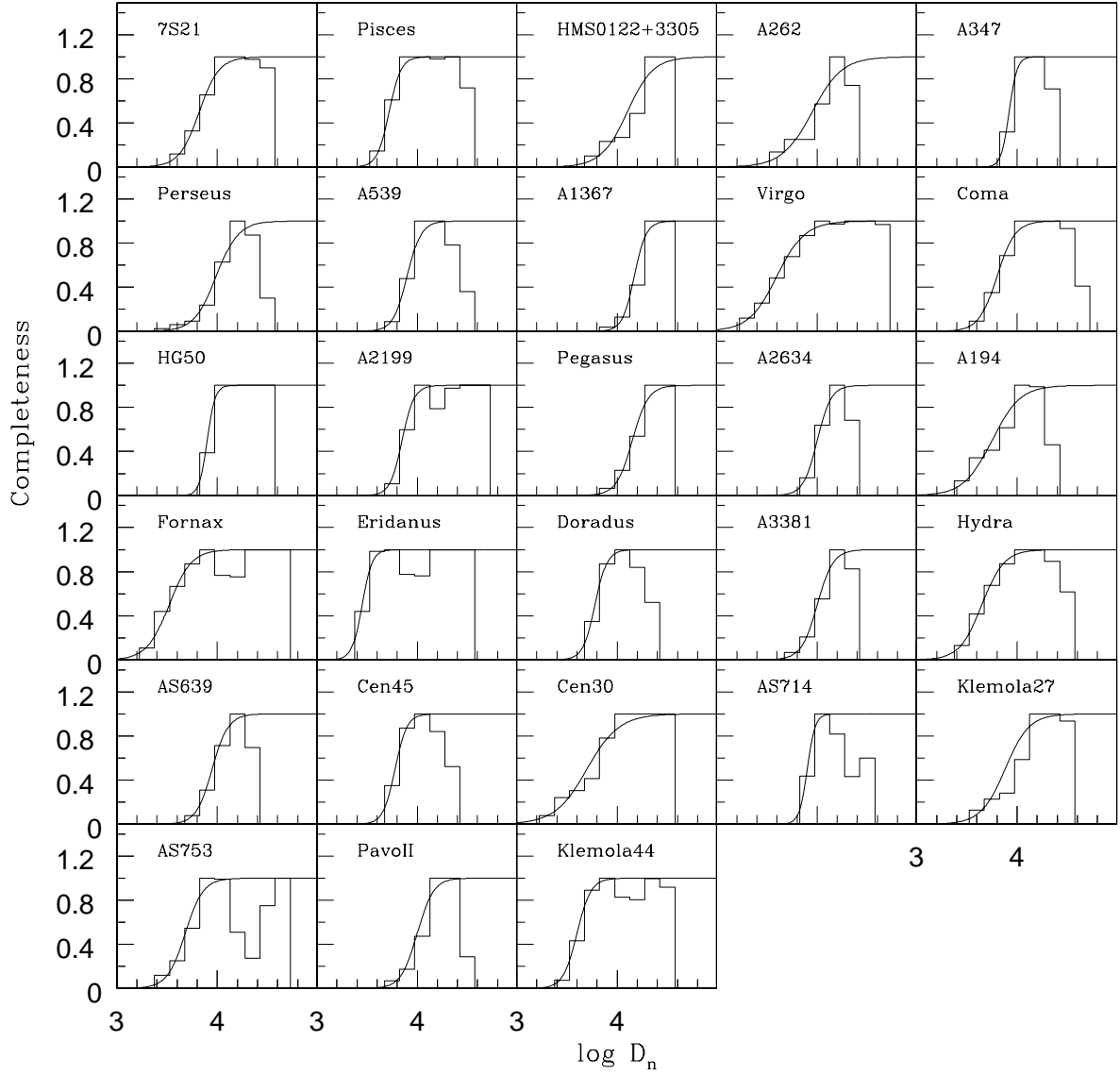,height=15truecm,bbllx=2truecm,bblly=5truecm,bburx=19truecm,bbury=23truecm }} 
\caption{Panels show the selection function for each cluster, 
computed from the ratio of the number of objects observed in the 
cluster and the number predicted by the fitted diameter distribution 
function. Solid curves show fits to the histograms of the form 
given by Equation~(\ref{eq:select}).}
\label{fig:sel_func_compl}
\end{figure}

Figure~\ref{fig:sel_func_compl} shows the completeness function
$C(D_n)$ for each cluster computed from the ratio between the number 
of objects observed in the cluster to the number predicted by the fitted
diameter-distribution function (equation~\ref{eq:diamfunc}).

In practice, this is done after binning the data in 
$\Delta y \equiv \Delta \log{D_n} = 0.2$ bins, and then smoothing 
with a Hanning filter (convolving with a [0.25, 0.50, 0.25] function)
to reduce the effects of small number statistics.  The solid
curve is a fit to the histograms using the function (G97)
\begin{equation} 
 C(y)= {1\over 1 + {\rm e}^{(y-y_f)/\eta}} 
\label{eq:select}  
\end{equation}  
to represent the completeness function, thereby further reducing the 
effects of small number statistics. Table~\ref{tab:param} gives
the parameters $y_f$ and $\eta$ of the completeness function
for each cluster.  At the bright end (large values of $y$) the 
completeness was normalized to unity, based on the fact that in all
clusters the brightest galaxies are always included in the cluster
sample. Comparison between the predicted and observed
diameter-functions for the nearby Virgo cluster are in good agreement
down to small values of $D_n$, indicating that we could have used Virgo
to estimate the completeness of the other clusters.

\begin{figure}
\centering
\mbox{\psfig{figure=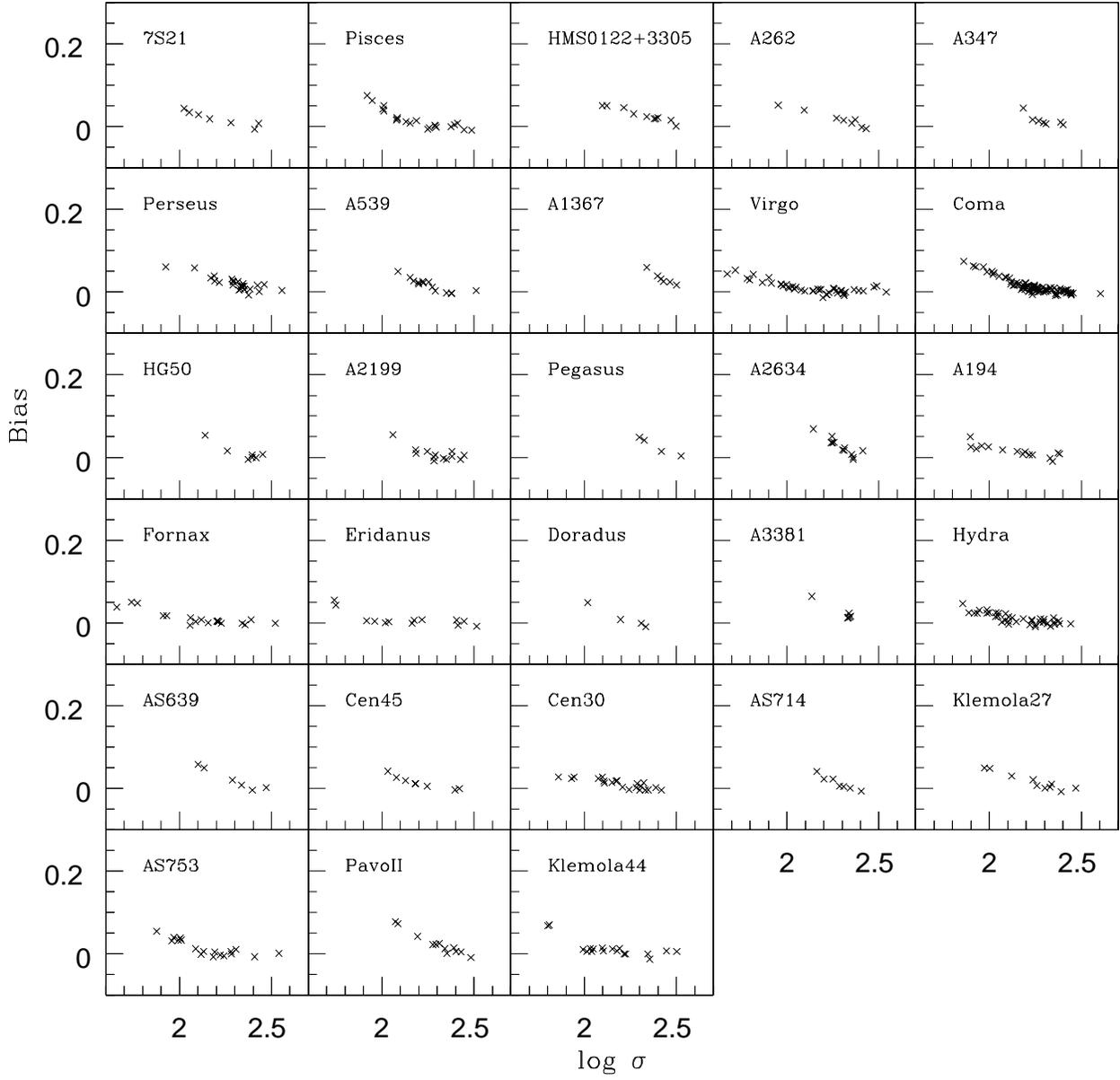,height=15truecm,bbllx=2truecm,bblly=5truecm,bburx=19truecm,bbury=22truecm }}
\caption{The incompleteness bias corrections that were applied to the individual measurements.}
\label{fig:DS_group_bias}
\end{figure}

Using this function as input, we estimated the bias correction
$B_{ik}$ for the $i$-th galaxy in the $k$-th cluster.  The results
after the final iteration are shown in Figure~\ref{fig:DS_group_bias}.  
For nearby clusters, such as Virgo and Fornax, the incompleteness  bias
correction is small, as expected.  For more distant clusters the
correction can be significant, with $\Delta y \sim 0.1$
(corresponding to $\Delta m \sim 0.5$ mag).

After the iterating, final values for the distance relation
coefficients are determined. Applying the condition that ``distant''
clusters, i.e., clusters beyond $3000$ kms$^{-1}$ (with a mean redshift
of $6000$ kms$^{-1}$), are at rest with respect to the CMB, and 
excluding clusters with suspisciously large peculiar velocities 
(see discussion below), we obtain the following final relation: 
\begin{equation} 
\label{eq:dnsigdir} 
 \log D_n = 1.203(\pm 0.023)\ \log\sigma_0 \ +\ 1.406(\pm 0.021), 
\end{equation}
where the error of the slope is derived by bootstrap re-sampling.  
The bootstrap error is based on the distribution of the slopes derived from
a large number of data sets constructed through random sampling of the
observed data set. The derived zero-point is consistent with the
value obtained by using the distance of Virgo as determined by the
Cepheid period-luminosity relation (Kelson \etal 2000). It is also
consistent with the value obtained by assuming that Coma is at rest with
respect to the CMB (the $D_n-\sigma$ relation given above leads to a
peculiar velocity of $\sim -72 \pm 189$ kms$^{-1}$ for Coma).

The error in the zero-point has two sources. The first is related to
the scatter in the distance relation and the procedure adopted in the
construction of the template relation.  This was estimated as
follows. We constructed data sets by randomly removing some points and
replacing them with others from the observed data set. For each
cluster the same fraction of data points were replaced, typically from
5\% to 25\%.  For clusters with few members ($\lsim 10$) for which
this was not possible due to the small number of cluster members, we
left out one or two observations in sequence.  We fixed the slope of
the $D_n-\sigma$ relation, and derived the zero-point from each
simulated data set. The random uncertainty in the zero-point is given
by the standard deviation of a Gaussian fit to the distribution of
these zero-points.  This yields an error of 0.018, corresponding to an
error of $\sim 4$\% in distance.  The second contribution to the
zero-point error is the uncertainty in the mean velocity of the
distant cluster sample used to set the zero-point of the
relation. This uncertainty is due to the finite number of clusters
used to sample the peculiar velocity field of the clusters. It is also
susceptible to cosmic variance.  The uncertainty in the mean peculiar
velocity of the cluster sample is given by $\sigma/\sqrt{N}$, where
$\sigma$ is the $rms$ of the clusters' peculiar velocity distribution
and $N$ is the number of clusters. Note, however, that since the
clusters in our sample are not randomly distributed the actual number
of clusters with uncorrelated velocities should be smaller than the 28
clusters considered. Here we estimate the error to be $\sim 400/4=100$
\kms, which corresponds to $100/4500\sim$ 2\% at the median distance
of the clusters in our sample. Adding in quadrature the different
contributions to the error budget we estimate the final error in the
zero-point to be $0.021$, with the main contribution coming from the
uncertainties associated with the distance relation.

\begin{figure}
\centering
\mbox{\psfig{figure=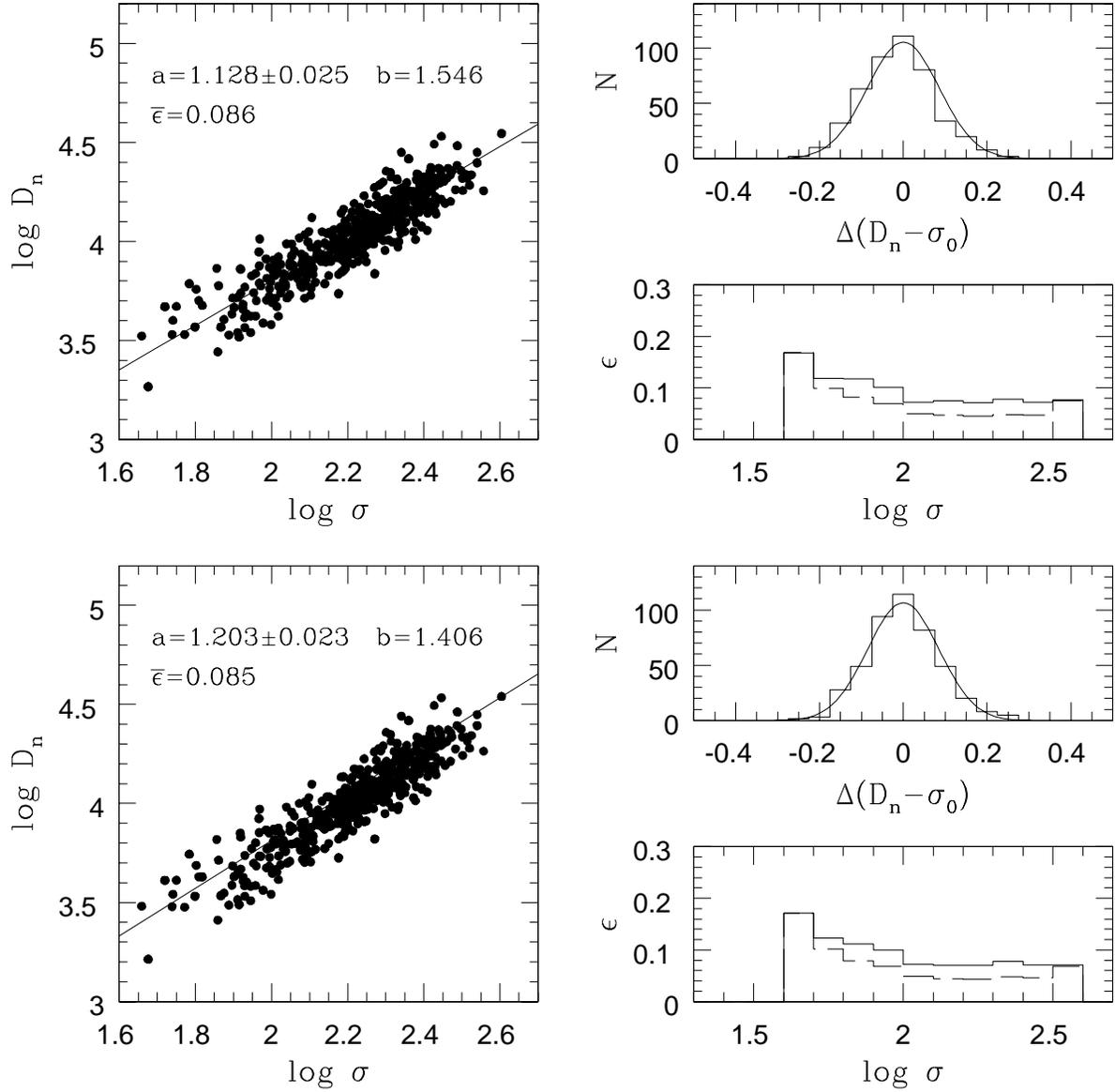,height=15truecm,bbllx=2truecm,bblly=5truecm,bburx=19truecm,bbury=23truecm }}
\caption{Panels
on the left show measurements before the bias 
correction is applied (upper), and the final corrected values 
derived from the iterative process (lower) as a function of 
$\sigma$. The line shows the derived distance relation. 
The values of the slope ($a$), zero-point ($b$), and the mean $rms$ 
scatter ($\bar{\epsilon}$) are also shown. 
Panels on the right show the distribution of the residuals 
relative to the $D_n-\sigma$ relation, as well as the distribution 
of the corresponding observed scatter (solid line) and intrinsic 
scatter (dashed line) as a function of $\sigma$.}
\label{fig:fit_dir_cor}
\end{figure}

Figure~\ref{fig:fit_dir_cor} shows the initial and final estimates of 
the distance relation together with the distribution of the observed
$rms$ scatter (solid histogram), and the intrinsic scatter (dashed 
histogram), as a function of $\sigma$. The intrinsic scatter was 
derived by subtracting the measurement uncertainties in quadrature 
from the $rms$ scatter of the fit---although it increases at 
low $\sigma$, its mean value is $\sim 0.06$~dex.  
This scatter may reflect differences in the stellar populations 
of the cluster member galaxies.  This possibility will be discussed 
in Section~\ref{stelpop}.  
The intrinsic scatter limits the accuracy of the derived distances.  
The mean of the total scatter $\bar{\epsilon}\sim$0.085 dex, 
yields a distance error $\Delta \sim 20\%$.  
This is comparable to the errors obtained using FP relations 
(\eg Hudson \etal 1997). 

We used the Coma cluster to test if our correction based on 
Monte-Carlo simulations is reliable.  For this cluster we extracted
sub-samples using different magnitude-limits and computed the
$D_n-\sigma$ relation for each individual sub-sample as follows: a)
without applying the bias correction described in Section~\ref{monte},
and b) correcting the slope for selection effects.  We found that
imposing a magnitude-limit biases the slope to lower values, but that
the Monte-Carlo technique used recovers the correct value of the slope
in each sub-sample.  Using galaxies in Coma we also checked whether
adopting different lower-limits in velocity dispersion biases the
slope of the distance relation. We found that increasing the value of
the velocity dispersion cutoff, for instance from 70 to 100 kms$^{-1}$
does not affect significantly our results, with the slope varying by
less than $2\%$ and the scatter remaining unchanged.

\begin{figure} 
\centering 
\mbox{\psfig{figure=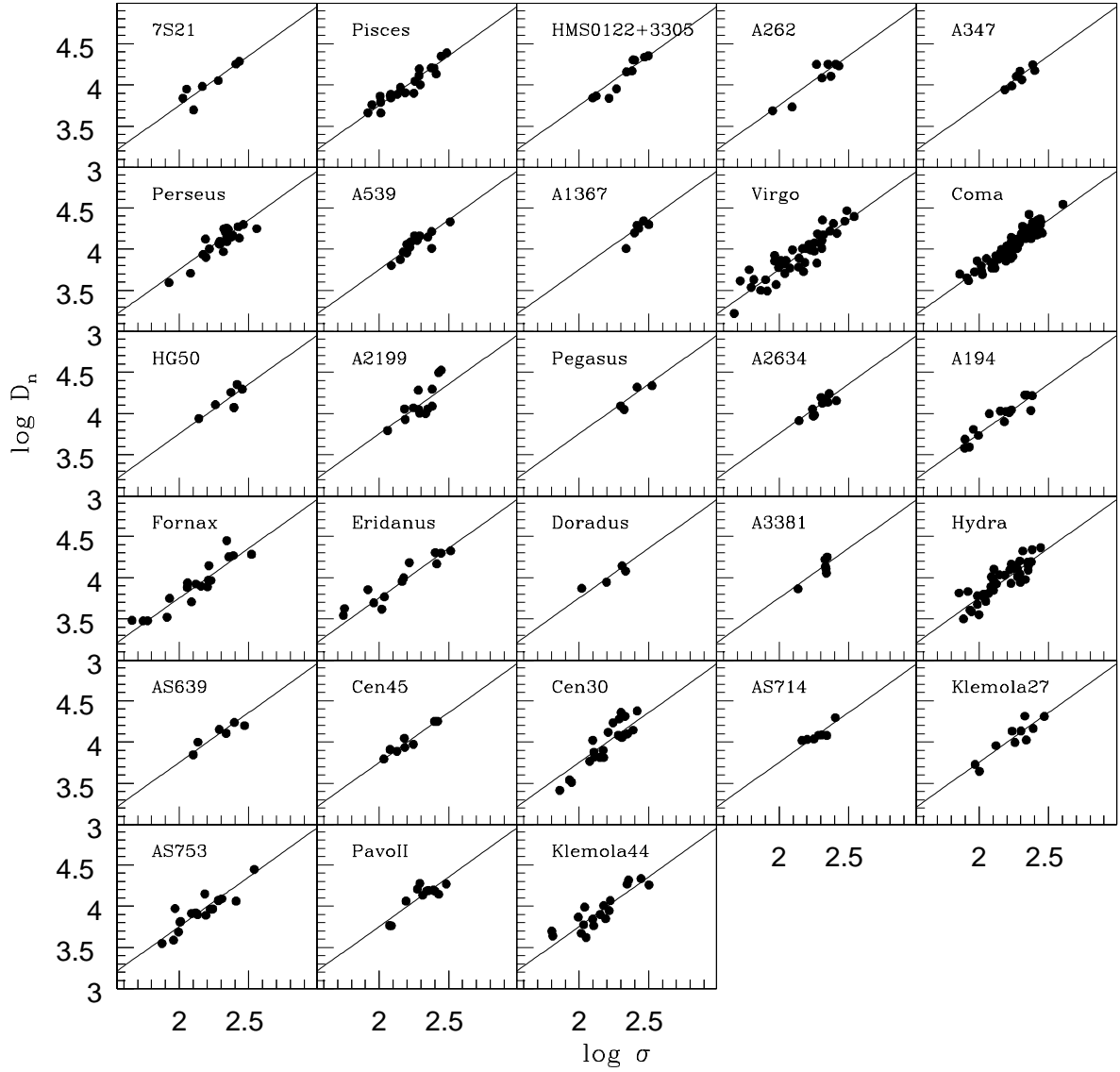,height=15truecm,bbllx=2truecm,bblly=5truecm,bburx=19truecm,bbury=23truecm }} 
\caption{Bias-corrected $D_n$ of each cluster member galaxy versus 
its velocity dispersion. Solid line shows the derived distance 
relation (Equation~(\ref{eq:dnsigdir})).}
\label{fig:DS_group_cor} 
\end{figure}

Figure~\ref{fig:DS_group_cor} shows the bias-corrected data points 
for all clusters and the final fit. A number of interesting cases 
are evident. For instance, HMS0122+3305, A2199, Cen30, exhibit clear
evidence of either spatial sub-structure or distinct galaxy
populations, and the galaxies in the cluster AS714 do not strictly
follow the template relation. The individual \dnsig exhibit a tilt
relative to the template relation.  We will return to these points
below.

To evaluate the robustness of our results, we derived the $D_n-\sigma$
relation when specific sub-samples of galaxies were excluded. The results 
are summarized in Table~\ref{tab:tests}: column (1) gives the sub-sample 
of galaxies removed (A $=$ ``peripheral'' objects defined in B01; 
B $=$ clusters whose individual $D_n-\sigma$ relations differ significantly 
($\Delta$ slope $\gsim$ 0.2) from equation~(\ref{eq:dnsigdir});  and 
individual clusters); column (2) the number of remaining galaxies which were 
used to compute the $D_n-\sigma$ relation; column (3), (4) and (5) the slope,
the zero-point, and the $rms$ scatter of the $D_n-\sigma$ relation obtained 
using the number of galaxies given in column (2).
Based on these tests we conclude that the variation of the slope, 
$a$, agrees with the formal error computed from the bootstrap 
re-sampling ($\sigma_a~\sim~0.023$).

\begin{figure}
\centering
\mbox{\psfig{figure=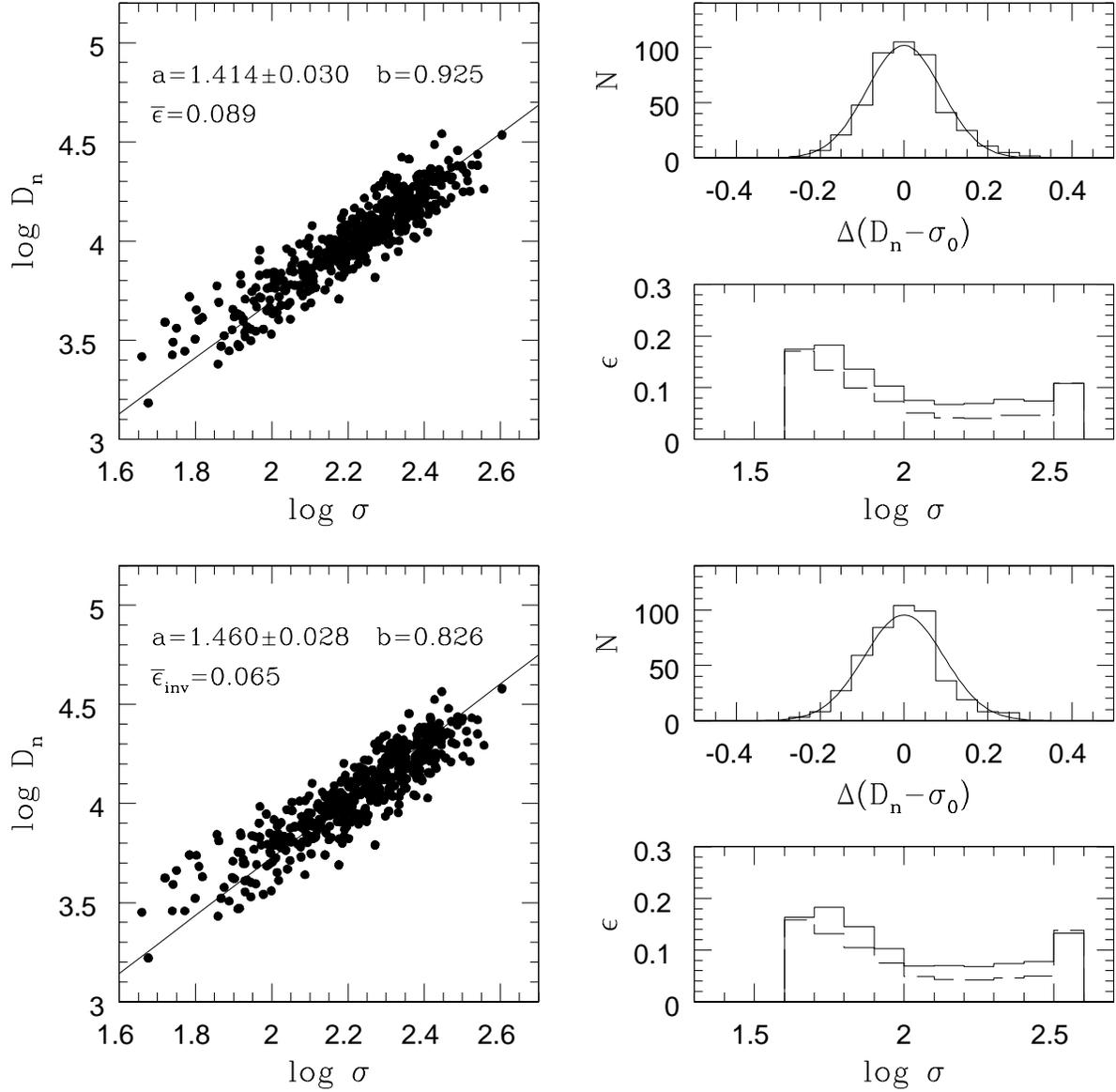,height=15truecm,bbllx=2truecm,bblly=5truecm,bburx=19truecm,bbury=23truecm }} 
\caption{Symbols in
the panels on the left show the bias corrected 
measurements; straight lines show the derived bivariate distance 
relation (upper) and the inverse relation (lower).  The values
of the slope ($a$), zero-point ($b$), and the mean 
$rms$ scatter ($\bar{\epsilon}$) are also shown. 
Panels on the right show the distribution of residuals relative to 
the $D_n-\sigma$ relation, together with the distribution of the 
corresponding observed scatter (solid line) and intrinsic scatter 
(dashed line), as a function of $\sigma$.}
\label{fig:fitbivinv}
\end{figure}

We have also computed the direct relation using orthogonal fits,
allowing for errors in both $\log{D_n}$ and $\log{\sigma}$, and for 
the inverse relation, ignoring the bias correction. The results are 
shown in the upper and lower panels of Figure~\ref{fig:fitbivinv}. 
The corresponding coefficients and scatter in $\log{D_n}$ are given 
in Table~\ref{tab:fits}.  The inverse relation is insensitive 
to the photometric selection and is, in principle, bias-free
if no a priori cut is made in that variable (see Strauss \& Willick
1995).  However, as some data from the literature in our sample 
are limited to galaxies with $\sigma\gsim 100$ kms$^{-1}$,
this assumption may not hold and, perhaps, in these cases treatement 
similar to that carried out for the distance-dependent parameter 
should be considered.

\begin{figure}[t!]
\centering
\mbox{\psfig{figure=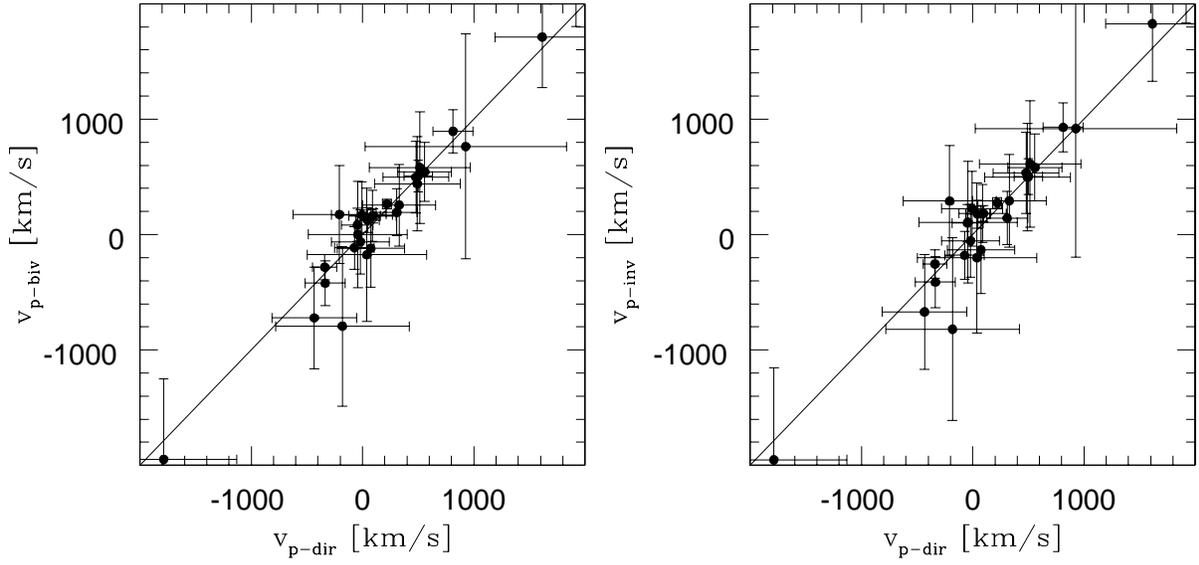,height=12truecm,bbllx=2truecm,bblly=13truecm,bburx=19truecm,bbury=27.5truecm }}
\caption{Cluster
peculiar velocities obtained using the direct
$D_n-\sigma$ relation (Equation~(\ref{eq:dnsigdir})) versus the values
computed using the bivariate (left) and the inverse relations 
(right).}
\label{fig:velpec_comp6}
\end{figure}

The three fitting relations in Table~\ref{tab:fits} clearly show that 
it is crucial to use a self-consistent fitting algorithm, to have a 
large and homogeneous set of data, and to correct for selection
biases.  For this analysis we found that even though the slopes 
and zero-points of the direct, bivariate, and inverse relations 
differ by more than $2\sigma$, the distances of the 28 clusters in 
the ENEARc sample agree well (see Section~\ref{vpec} and 
Figure~\ref{fig:velpec_comp6}).

The coefficients determined from our sample are compared with those 
found by other authors in Table~\ref{tab:tablit}.  Our results 
are generally in good agreement with previous determinations, except 
for those of Baggley (1996) and Lucey \etal (1997). 
Saglia \etal (2001) have recently revised  Baggley's result giving 
both a slope and zero-point comparable to our values. 
Lucey \etal used two distant clusters, A2199 and A2634, to obtain 
their results. Our analysis shows that A2199 has an individual 
$D_n-\sigma$ relation which differs significantly from 
equation~(\ref{eq:dnsigdir}) (see Table~\ref{tab:clusindiv}), 
while A2634 has a high peculiar motion (see next section for more 
details).

\section{Cluster Peculiar Velocities}
\label{vpec}

\begin{figure}
\centering
\mbox{\psfig{figure=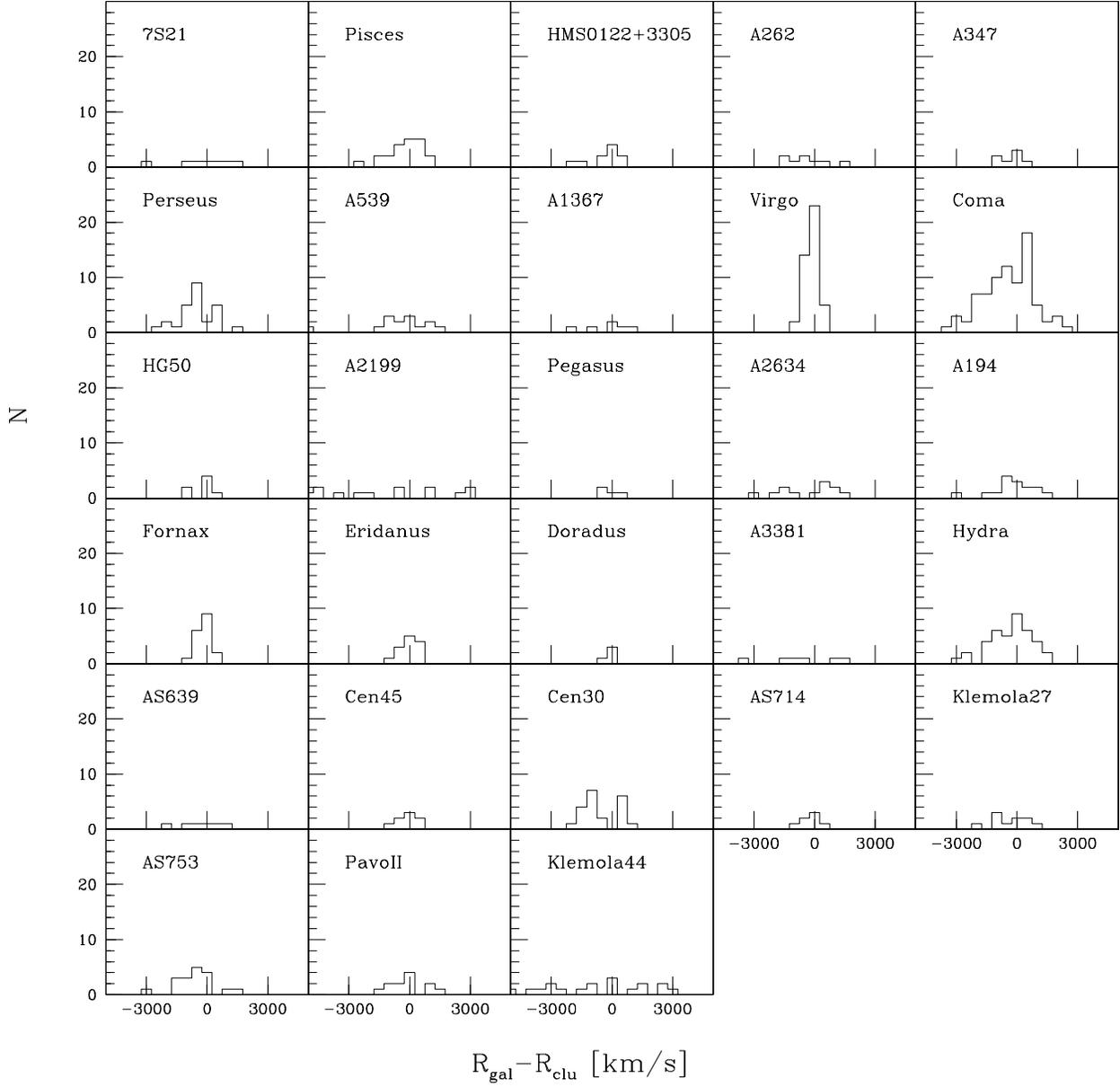,height=15truecm,bbllx=2truecm,bblly=5truecm,bburx=19truecm,bbury=22truecm }}
\caption{The distribution of 
the difference between the individual 
galaxy distances derived from Equation(~\ref{eq:dnsigdir}), 
and the error-weighted mean of the distribution, which is used 
as the cluster distance.}
\label{fig:DS_group_dist}
\end{figure}

We compute distances to galaxies in clusters using the ``direct''
template relation found in the previous section. 
Figure~\ref{fig:DS_group_dist} shows the differences in distance between
each individual galaxy and its cluster, the distance of which was
computed as the error-weighted mean of the galaxy distances in 
the cluster. For the best sampled clusters (\eg Virgo, Fornax) the 
distance distributions have well defined peaks and small scatter,
resulting in good mean distances.  However, there are a few complex 
cases where clusters exhibit sub-structure (\eg HMS0122+3305, 
Perseus, Coma, and Cen30). Also, there are clusters which either 
show large scatter and poorly defined peaks (\eg A2199, A2634, and
Klemola44).  And finally, there are clusters which have only a few 
galaxies; typically, these are either nearby small groups (\eg 7S21, 
A347, A1367, HG50, Pegasus, Doradus, AS714) or very distant clusters 
(\eg A3381) with large distance uncertainties.

The cluster distances were corrected for homogeneous Malmquist bias 
(following Lynden-Bell \etal 1988, the estimated distance is 
multiplied by $\exp(3.5\bar{\epsilon}^2/N_g)$, where $N_g$ is the 
number of cluster galaxies);  this correction generally amounts to 
less than $\sim 3\%$  of the distance for the smallest groups.

The radial component of the peculiar velocity of each cluster, 
$v_p=cz_{\rm{cor}}-R$, was computed using the Malmquist corrected 
distance $R$, and the mean cluster radial velocity $cz$ presented in 
B01 and corrected for the cosmological effect: 
\begin{equation}
cz_{\rm{cor}} = 
 cz - \log \left (\frac{1+(7/4) (cz/c)}{1+(7/4) (cz_{\rm Coma}/c)},
\right)
\label{eq:czcorr}
\end{equation} 
where $cz_{\rm{Coma}}$ is the Coma cluster radial velocity and
$c$ is the speed of light (Lynden-Bell \etal 1988).  
Baggley (1996) computed a more accurate cosmological correction, but 
for nearby galaxies equation~(\ref{eq:czcorr}) is a good approximation.

The measured cluster distances and peculiar velocities are presented in
Table~\ref{tab:clustervpec}: column (1) gives the cluster name; column
(2) the number of observed cluster galaxies; columns (3) and (4) 
are the cluster's  Galactic coordinates; column (5) its
redshift determined from the group finding algorithm (see
B01) and its error in the CMB frame; column (6) is the computed cluster
distance and its error; and column (7) gives the cluster peculiar
velocity and its error in the CMB frame.

\begin{figure}
\centering
\mbox{\psfig{figure=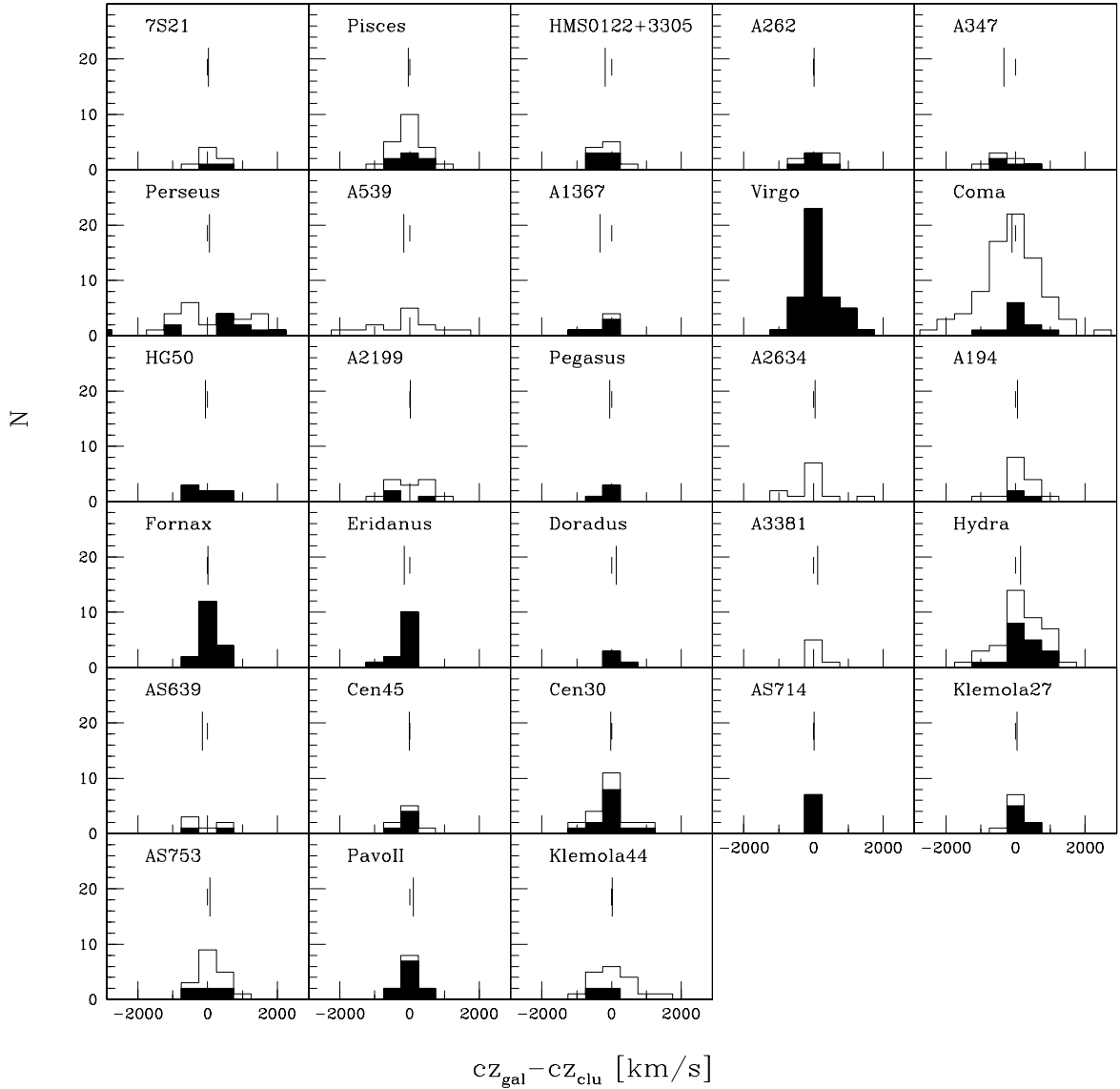,height=15truecm,bbllx=2truecm,bblly=5truecm,bburx=19truecm,bbury=23truecm }} 
\caption{The distribution of differences between the galaxy redshifts 
and that of the cluster to which they were assigned. 
Open histograms show this distribution for all the galaxies in the 
cluster, regardless of their aparent magnitude. Solid histograms show 
the distribution for only those galaxies which were identified as 
members of the parent cluster by applying an objective group-finding 
algorithm to complete, magnitude-limited redshift surveys.}
\label{fig:DS_group_cz}
\end{figure}

How the sample of galaxies in a cluster is chosen can lead to 
significant differences in the determined mean velocity. 
Figure~\ref{fig:DS_group_cz} shows the galaxy redshift distribution 
in each cluster.  Open histograms show the distribution of 
differences in redshift between the individual galaxies and the 
redshift assigned to the cluster. Solid histograms show this 
distribution for galaxies which were selected by applying the 
group-finding algorithm to complete but magnitude-limited redshift 
surveys (see Section 2). 
The figure shows that the fraction of galaxies in some clusters 
identified by the algorithm is significantly smaller than the total 
number of galaxies used in this paper (\eg Coma).  
For such clusters, using all galaxies (open histograms) or using 
only this subset (filled histograms) may provide different estimates 
of the cluster's mean redshift, velocity dispersion, and other 
parameters.  Nevertheless, the figure suggests that the mean redshift 
remains about the same, even though the solid histograms are likely 
to underestimate the dispersion around the mean redshift.  

The error-weighted mean cluster redshift of early-type galaxies only
(long vertical line) and that given by the group finding algorithm
(short vertical line) are also shown in Figure~\ref{fig:DS_group_cz}.
For the latter, the sample of galaxies assigned to a group/cluster
included all morphological types. Note the significant redshift
differences, occasionally as large as 300 kms$^{-1}$.  The most
deviant cases ($\gsim 2\sigma$, where $\sigma$ is the error in the
mean cluster redshift of early-type galaxies -- i.e., the error on the
position of the long vertical line) are A347, A539, A1367, Eridanus,
Doradus, and Pavo~II.  This suggests that using a sub-sample of
galaxies in a cluster (especially when only few objects are selected)
to compute the cluster redshift may introduce an error which can, in
some cases, be large.  This possibility has been ignored in the past,
and may account for some disparities in the measurements of the
peculiar velocity.

\begin{figure}
\centering
\mbox{\psfig{figure=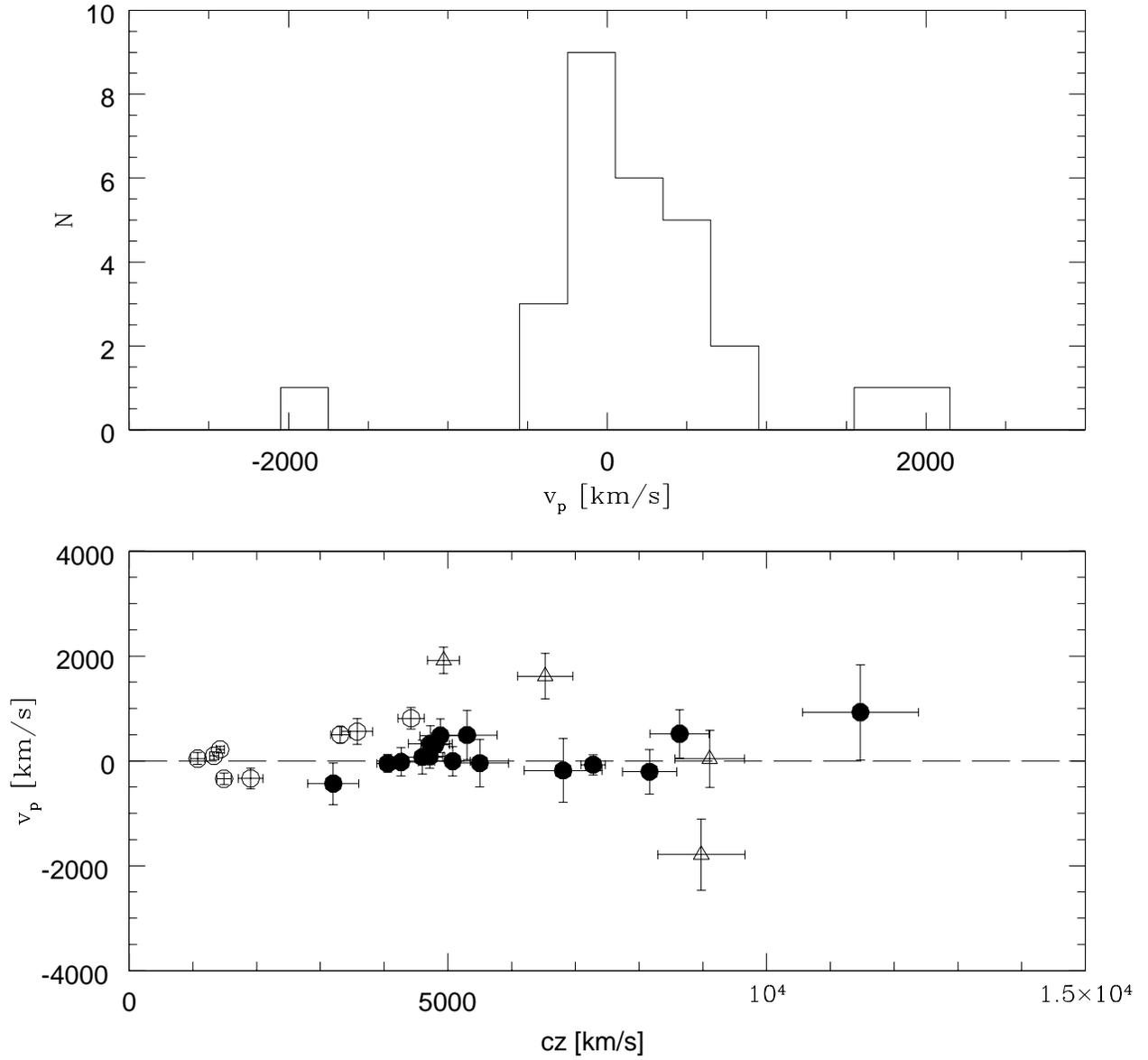,height=15truecm,bbllx=2truecm,bblly=5truecm,bburx=19truecm,bbury=23truecm }}
\caption{Distribution of
cluster peculiar velocities (upper panel) and cluster peculiar 
velocities versus the estimated distances (lower panel).
Filled circles represent the ``distant'' clusters used for the final 
calibration of the $D_n-\sigma$ relation; open circles represent 
either nearby clusters or clusters which have suspiciously large peculiar
velocities; and open triangles indicate clusters that were not observed 
by our survey.}  
\label{fig:velpec_hist}
\end{figure}

The upper panel of Figure~\ref{fig:velpec_hist} shows the distribution
of cluster peculiar velocities.  The lower panel shows that 
velocities do not depend on the estimated distances; 
large peculiar velocities occur both at small and large distances.  
Filled symbols are for the ``distant'' clusters used in the final 
calibration of the $D_n-\sigma$ relation---the subsample which is 
required to be at rest relative to the CMB. 
Open circles indicate nearby clusters plus three
additional clusters in the Great Attractor (\eg Lynden-Bell et al. 1988) 
region:  Cen30, AS714, and AS753.  Triangles show clusters with data 
exclusively from the literature. The $1\sigma$ error bars were
computed by adding the distance and the cluster mean redshift errors 
in quadrature. The errors in distances were taken to be 
$\Delta/\sqrt{N}$, where $\Delta$ is the fractional distance error 
derived from the scatter of the composite distance relation, and $N$ 
is the number of galaxies observed in the cluster. The error in the 
cluster redshift is estimated as $\sigma_{\rm cl}/\sqrt{N^\prime}$, 
where $\sigma_{\rm cl}$ is the velocity dispersion of the cluster and
$N^\prime$ is the number of galaxies in the group catalog.

The distribution shown in the upper panel of 
Figure~\ref{fig:velpec_hist}, which includes all 28 clusters, 
has an error-weighted mean of $151 \pm 75$ kms$^{-1}$, with a scatter 
of $399\pm 73$ kms$^{-1}$.  The bottom panel shows that there are 
three obvious outliers: A2634, AS639, and Cen45, all based on data 
from the literature.  Other clusters with large ($>2\sigma$) 
peculiar velocities are Cen30 ($500 \pm 153 $ kms$^{-1}$), 
AS714 ($ 559 \pm 245 $ kms$^{-1}$), 
and AS753 ($812 \pm 204 $kms$^{-1}$);
all are located near the Great Attractor. 
If these clusters are removed from the sample,
the mean peculiar velocity of the remaining 22 clusters is $71 \pm 51$
and the $rms$ one-dimensional cluster velocity is $239 \pm 46$
kms$^{-1}$.  This is comparable to what is measured from the SCI
sample (G97) $266\pm30$ kms$^{-1}$ (Giovanelli 1998).  
This small one-dimensional $rms$
cluster velocity has important implications for cosmological
parameters (\eg Giovanelli 1998; Borgani \etal 2000a). 

Notes to additional problematical clusters can be found in 
Appendix~A. Most of these clusters appear to suffer from the effects 
of substructure; they have a history of discrepant peculiar motions 
reported in the literature.

\begin{figure}
\centering
\mbox{\psfig{figure=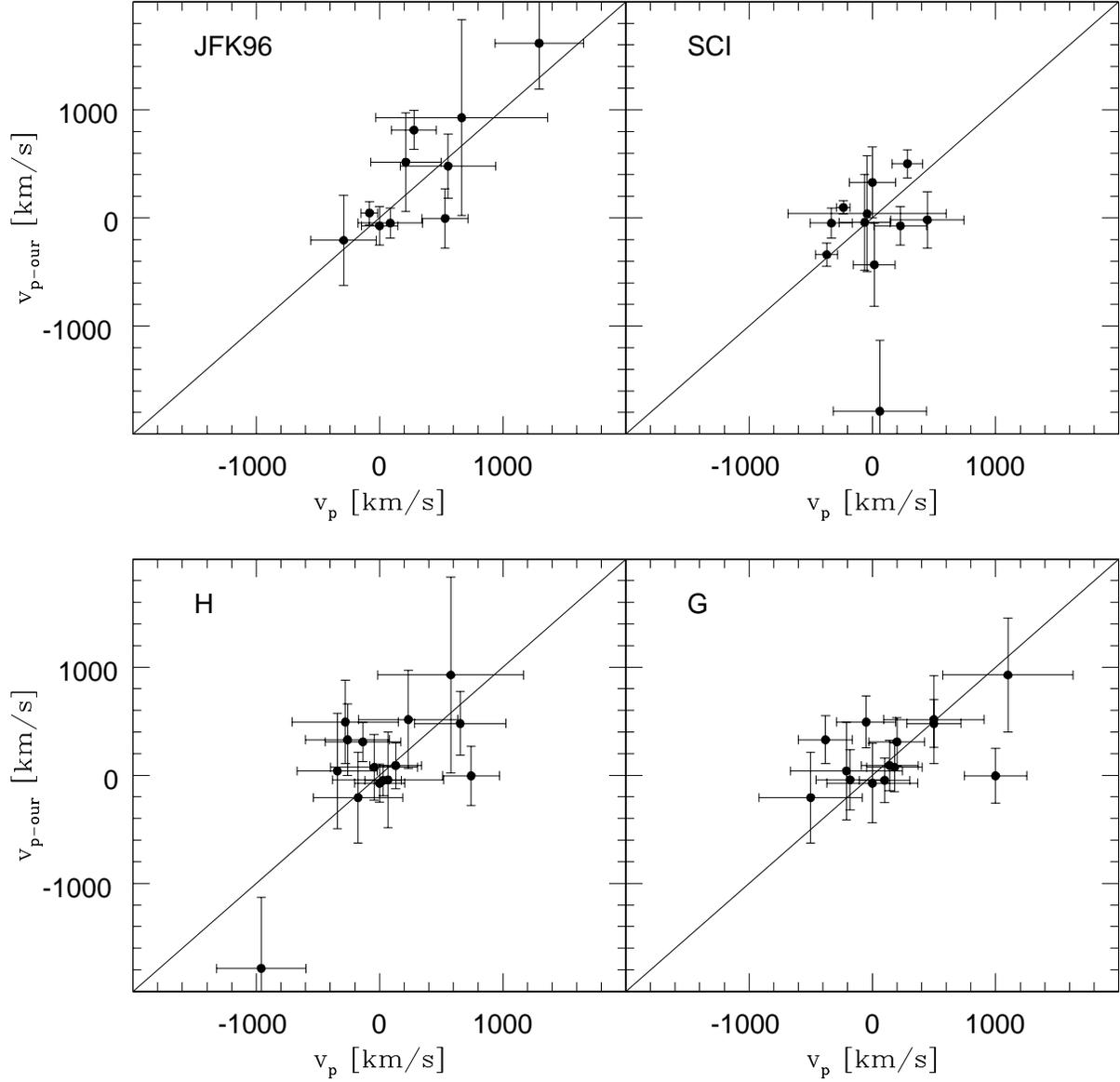,height=15truecm,bbllx=2truecm,bblly=5truecm,bburx=19truecm,bbury=23truecm }} 
\caption{Cluster peculiar velocities obtained using 
Equation~(\ref{eq:dnsigdir}) versus the values computed by
J{\o}rgensen et al. (1996) (JFK96), Giovanelli et al. (1997) (SCI), 
Hudson et al. (1997) (H), and Gibbons et al. (1998) (G).}
\label{fig:velpec_comp5}
\end{figure}

The large overlap between our cluster sample and the literature allows
a global comparison of the measured peculiar velocities. 
Figure~\ref{fig:velpec_comp5} shows our cluster peculiar velocities
($v_p$), computed using the direct \dnsig relation, for the clusters
we have in common with J{\o}rgensen \etal (1996) (10 clusters), SCI
(11 clusters), Hudson \etal (1997) (15 clusters), and Gibbons \etal
(2001) (15 clusters). This figure shows that except for 
A2634, A194, and AS753 our measurements of the cluster
peculiar velocities are in good agreement with those reported in
literature when the measurement errors are taken into account.
We find mean differences of $79\pm 91$ kms$^{-1}$ (J{\o}rgensen
\etal), $ 182 \pm 94 $ kms$^{-1}$ (SCI), $-9 \pm 96$ kms$^{-1}$
(Hudson \etal) and $ -53\pm 93$ kms$^{-1}$ (Gibbons \etal).  All
clusters are in the same rest frame to within $2\sigma$.  This
agreement shows consistency between different determinations
of cluster distances (e.g., those based on the \dnsig and/or FP
relation) and, more importantly, consistency with the TF relation for
spiral galaxies.  

Figure~\ref{fig:velpec_comp6} compares the peculiar velocities
computed using the bivariate relation, corrected for selection bias
(left panel), and the inverse relation (right panel) with
those determined using the direct relation.  The mean differences are
$-43 \pm 32$ kms$^{-1}$ with a scatter of 74 kms$^{-1}$, and $-58 \pm
38$ kms$^{-1}$ with a scatter of 89 kms$^{-1}$ for the bivariate and
inverse relations, respectively.
This shows that the peculiar velocities of the clusters are largely 
insensitive to the fitting procedure, whether the direct, bivariate 
or inverse relation is used.

\section {Dependence of the distance relation on galaxy properties}
\label{implications}

To use the composite \dnsig relation as a distance indicator, we 
should demonstrate that systematic cluster-to-cluster differences 
are small and that the computed cluster distances are unaffected by 
differences in the morphological mix of the galaxy population, 
different stellar populations or other cluster properties. 
Furthermore, the measured peculiar velocities must be free of any 
other systematic effects such as extinction, and contamination by
interlopers.  This can be tested  by examining the residuals from 
the distance relation which, for our data, exceed the estimated 
measurement errors of the \dnsig parameters.  
(Note that testing to see if the distance relation depends on whether 
or not the galaxies are in clusters or in less dense environments 
is not the subject of this paper.)  
All the tests below suggest that cluster-to-cluster variations 
are indeed small.

\subsection{Results for individual clusters}
\label{individual}

\begin{figure}
\centering
\mbox{\psfig{figure=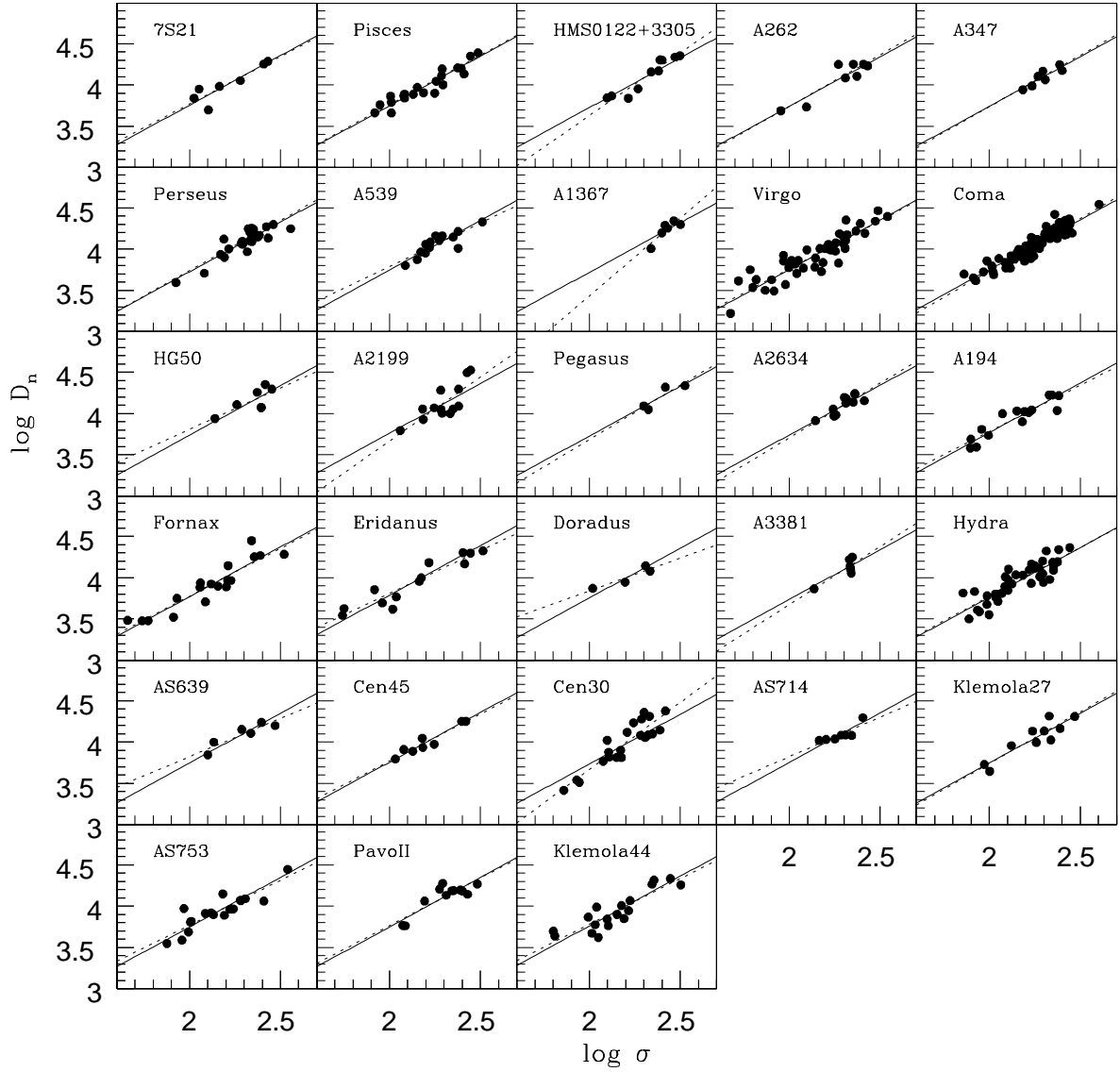,height=15truecm,bbllx=2truecm,bblly=5truecm,bburx=19truecm,bbury=23truecm }} 
\caption{The individual
cluster $D_n-\sigma$ relations obtained by 
fitting the bias-corrected data points of the cluster (dashed line).
The solid line in all panels shows the template
distance relation given by Equation~(\ref{eq:dnsigdir}).}
\label{fig:DS_group_indiv}
\end{figure}

Figure~\ref{fig:DS_group_indiv} plots the measured values of 
$D_n$ and $\sigma$ for the galaxies in each cluster along with 
the (incompleteness corrected) fit (dashed line) and the composite 
template relation (solid line).  The parameters for the individual 
fits are given in Table~\ref{tab:clusindiv}: 
column (1) gives the cluster name;
column (2) the number of cluster galaxies entering the \dnsig relation; 
column (3) the slope and its error; 
column (4) the mean scatter in $\log{D_n}$ of the data points 
relative to the individual fit; 
column (5) the zero-point offset between the individual and the
composite template relation (equation~(\ref{eq:dnsigdir})); 
column (6) the scatter relative to the fit obtained using the 
slope of the composite template relation but allowing the individual 
zero-point to vary;
column (7) the intrinsic scatter computed using the scatter listed in
column (6) and the errors of the measured parameters; and 
column (8) gives the fraction of the cluster galaxies in the observed 
sample that are ellipticals (F$_E$~=~$\rm{N_E}/(\rm{N_{E} + N_{S0}})$). 

Figure~\ref{fig:DS_group_indiv} shows that for most clusters the
individual fits have nearly the same slope as the template. The figure
also shows the benefit of combining all the data, because the slope
for the poorer systems in the sample is poorly determined.
Significant departures ($\Delta$ slope $\gsim 0.2$) are seen for
HMS0122+3305, A1367, HG50, A2199, Doradus, A3381, AS639, Cen30, and
AS714. The main cause for the tilt of an individual \dnsig is the 
small number of galaxies in the cluster. 
(For example, Table~\ref{tab:clusindiv} shows that the tilt of 
the individual \dnsig relations does not correlate with the fraction 
of ellipticals in the cluster, although some incorrect morphological 
classifications may still be present.)  Indeed, even a single 
galaxy can cause a significant deviation from 
equation~(\ref{eq:dnsigdir}).  As discussed above (see also 
Appendix~A), many of these clusters show large motions.  Also, 
recall that A2199 and Cen30 are parts of two-component systems 
(A2199-97 and Cen30-45).
  
In general, Table~\ref{tab:clusindiv} shows that the individual fit
does not improve the scatter significantly and that variations are 
likely due to poor statistics. The source of the intrinsic scatter 
is still not understood, though the largest contributions to it  
probably arise from intrinsic differences in the dynamical structures 
of the cluster galaxies rather than from errors in the photometry
and spectroscopy.  Whether these intrinsic differences produce 
systematic errors in the distance determination is unknown. 
However, because we treat the thickness of the relation as an 
uncertainty in the derived distance, the impact of such scatter 
should not alter our conclusions about large-scale motions in the 
universe (see also Section~\ref{stelpop}).

\begin{figure}
\centering
\mbox{\psfig{figure=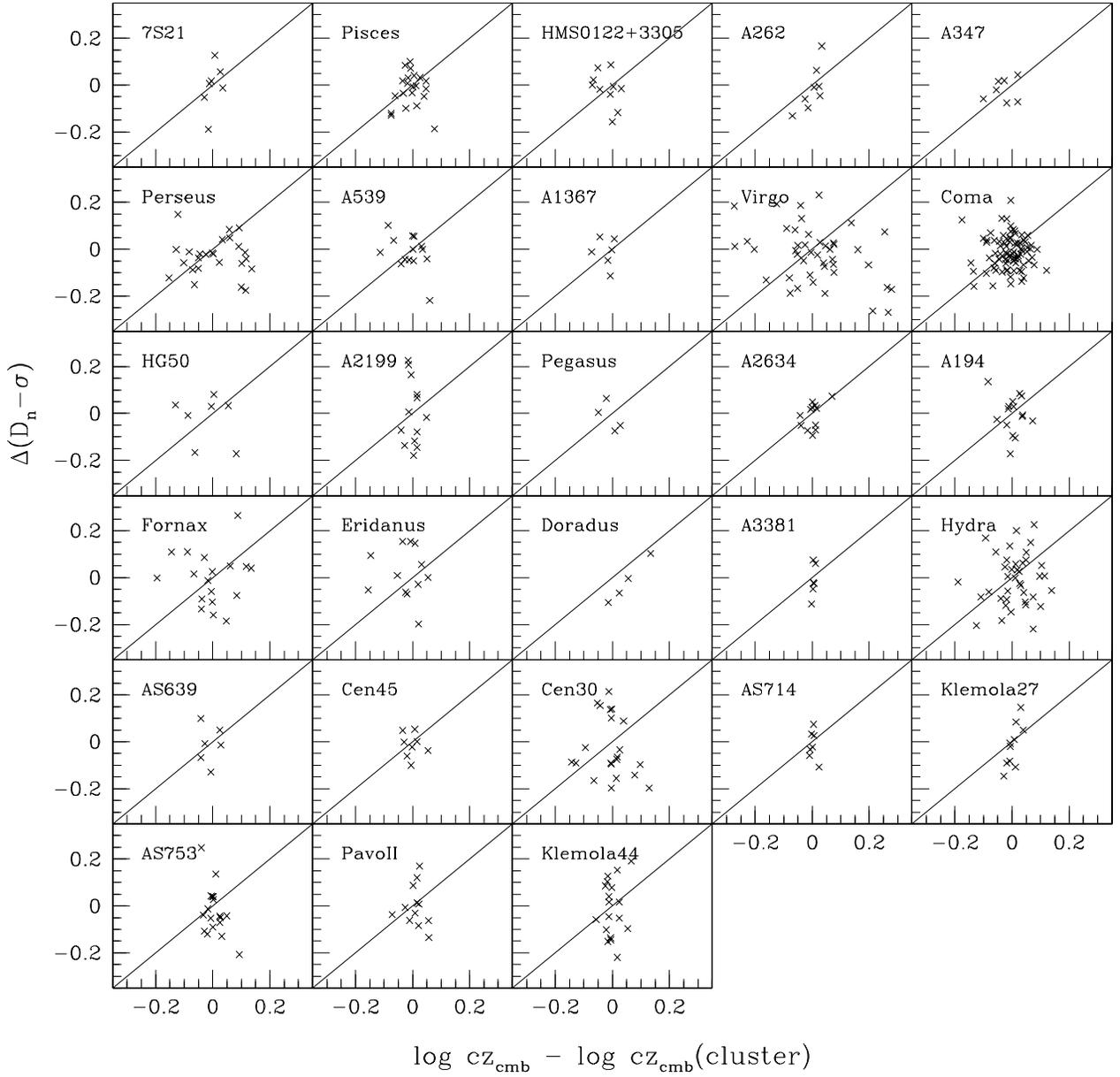,height=16truecm,bbllx=2truecm,bblly=5truecm,bburx=19truecm,bbury=23truecm }} 
\caption{The residual, relative to the 
distance relation, of each 
galaxy, as a function of the difference between the galaxy's 
redshift and that adopted for its parent cluster.}
\label{fig:DS_group_assign}
\end{figure}

Figure~\ref{fig:DS_group_assign} can be used to examine the impact 
of interlopers. It shows the residual of each galaxy from the 
distance relation as a function of the difference between the 
galaxy's redshift and that of the parent cluster. 
Field galaxies contaminating the sample would lie along the $45^\circ$ 
line shown in each panel---no such effect is seen.  
This is a consequence of our membership assignment and the fact that 
early-types are more likely to reside at the central regions of 
clusters.  

\begin{figure}[t!]
\centering
\mbox{\psfig{figure=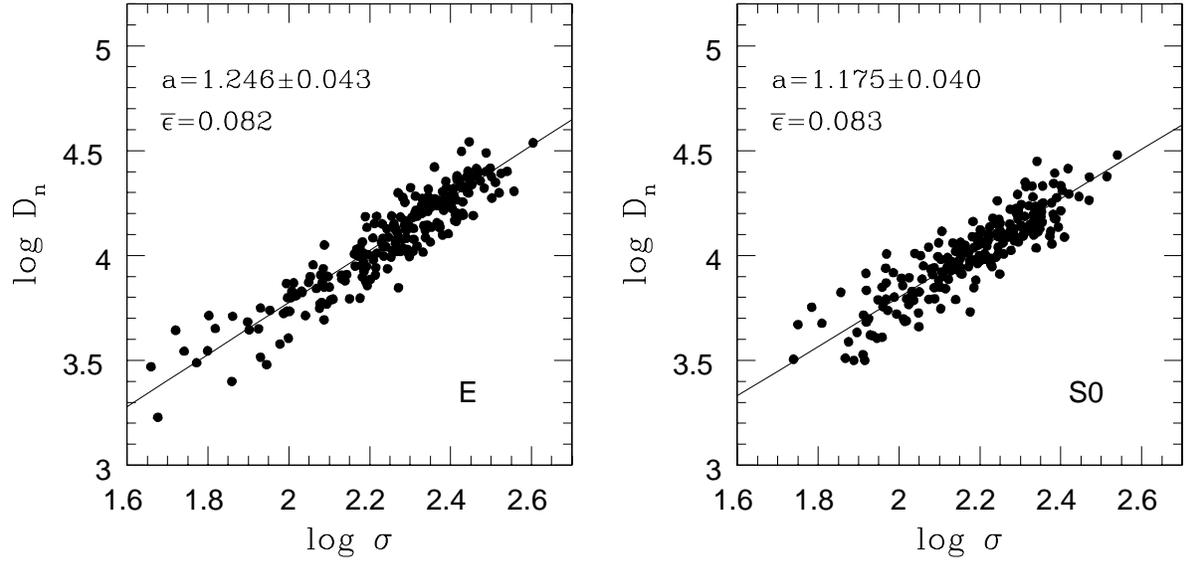,height=12truecm,bbllx=2truecm,bblly=13truecm,bburx=19truecm,bbury=27.5truecm }}
\caption{The $D_n-\sigma$ relation obtained 
from ellipticals (left panel) and S0s (right panel). 
The slope ($a$) of the relation and the corresponding mean 
$rms$ scatter ($\bar{\epsilon}$) are also shown.}
\label{fig:fit_dir_cor_E_S0}
\end{figure}

To study the effects of morphology, we split the sample into ellipticals (T
$\leq -3$) and S0s (T = -2), using Lauberts \& Valentijn's (1989)
classifications. First, we computed the $D_n - \sigma$ relation for the E and
S0 galaxies separately. Figure~\ref{fig:fit_dir_cor_E_S0}
shows these relations for the ellipticals (left panel) and S0s (right panel);
the difference in the slope of the two distance relations is $0.071\pm 0.059$,
which is significant at $< 2\sigma$ level, and the scatter is
comparable. Second, we determined the relative shifts which were required if a
linear relation of the same slope as of the composite template relation
($a=1.203$, see Equation~\ref{eq:dnsigdir}), was to fit the relation in each of
the subsamples. The difference in the intercept is not statistically
significant and the scatter is comparable. These results justify our neglect of
any morphological biases (Section~\ref{method}).

\begin{figure}[t!]
\centering
\mbox{\psfig{figure=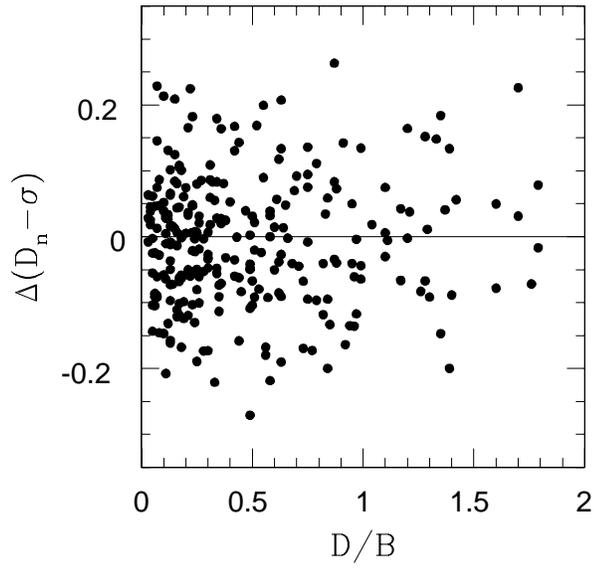,height=12truecm,bbllx=-2truecm,bblly=13truecm,bburx=19truecm,bbury=27.5truecm }}
\caption{The residuals, relative to the distance relation, as a
function of the galaxy $D/B$ ratio.}
\label{fig:residDB}
\end{figure}

To test further the above result one could, instead, 
consider the residuals in the $D_n-\sigma$ relation as a 
function of the $D/B$ ratio.  
Unfortunately, in practice, $D/B$ is available only for those 
galaxies in our sample which were observed by us; the data compiled 
from the literature used one component models to derive global 
photometric parameters. The 223 cluster galaxies for which we have 
our own photometric measurements show no correlation between the 
residuals of the $D_n-\sigma$ relation and the $D/B$ ratio 
(see Figure~\ref{fig:residDB}).  

\subsection{Stellar Populations}
\label{stelpop}

Earlier in this paper we found that the scatter of galaxies relative
to the template distance relation is roughly twice what can be
accounted for by measurement errors. The additional scatter has been
attributed, by several authors, to differences in stellar
populations. In the context of distance measurements, we must check if
these differences can lead to systematic errors in the distance, and
therefore to spurious peculiar velocities.

\begin{figure}
\centering
\mbox{\psfig{figure=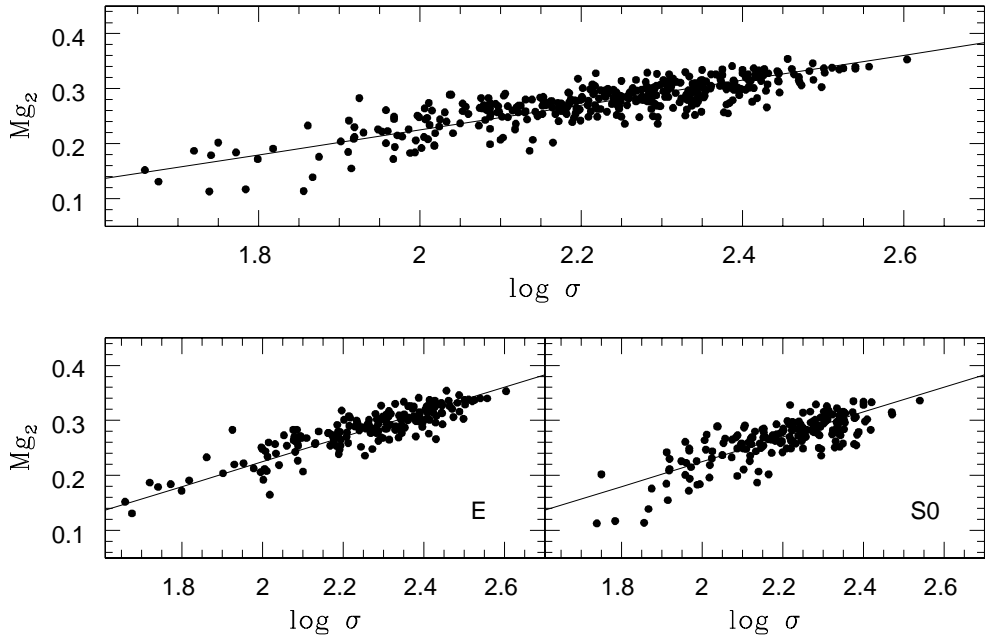,height=9truecm,bbllx=2truecm,bblly=12truecm,bburx=19truecm,bbury=25truecm }} 
\caption{(Upper panel) Measurements of the Mg$_2$ index versus the velocity dispersion
for the whole cluster sample. The solid line is the Mg$_2-\sigma$
relation derived from the bivariate fit.  (Lower panels) As in the
upper panel, but here the sample is split into ellipticals (left) and
S0s (right). In both lower panels, the solid line has the same
slope as in the upper panel (for the whole sample) while the
zero-point has changed.}
\label{fig:fit_mg2}
\end{figure}

To study the effects of different stellar populations, we use the
Mg$_2-\sigma$ relation, which is supposed to be distance independent.
It was computed for all galaxies in the sample after sorting them
according to their morphological types (see Figure~\ref{fig:fit_mg2}).
The parameters of the orthogonal fits to the Mg$_2-\sigma$ relation
are given in Table~\ref{tab:mg2fit}.  Column (1) of the table gives
the morphological types which are in the sample; column (2) the number
of galaxies; column (3) the slope computed from the whole sample and
its error; column (4) the zero-point and its error computed by fixing
the slope to the value reported in column (3); and column (5) the
scatter relative to the relation. Note that the coefficients
describing the linear fit obtained here differ slightly ($< 1\sigma$)
from those of Bernardi \etal (1998).  This is because we now include
Mg$_2$ measurements from other authors, scaling them to our
system. The table shows small differences in the zero-point between
ellipticals and S0s, although the S0 galaxies have a larger scatter
than ellipticals. This is partially due to the small number of
galaxies with low velocity dispersions in both sub-samples. Although
one expects S0 galaxies to form a less uniform class of objects than
ellipticals, the differences we find are small. Therefore, this
analysis suggests that the $D_n-\sigma$ relation does not depend on
differences in stellar populations.
  
\begin{figure}
\centering
\mbox{\psfig{figure=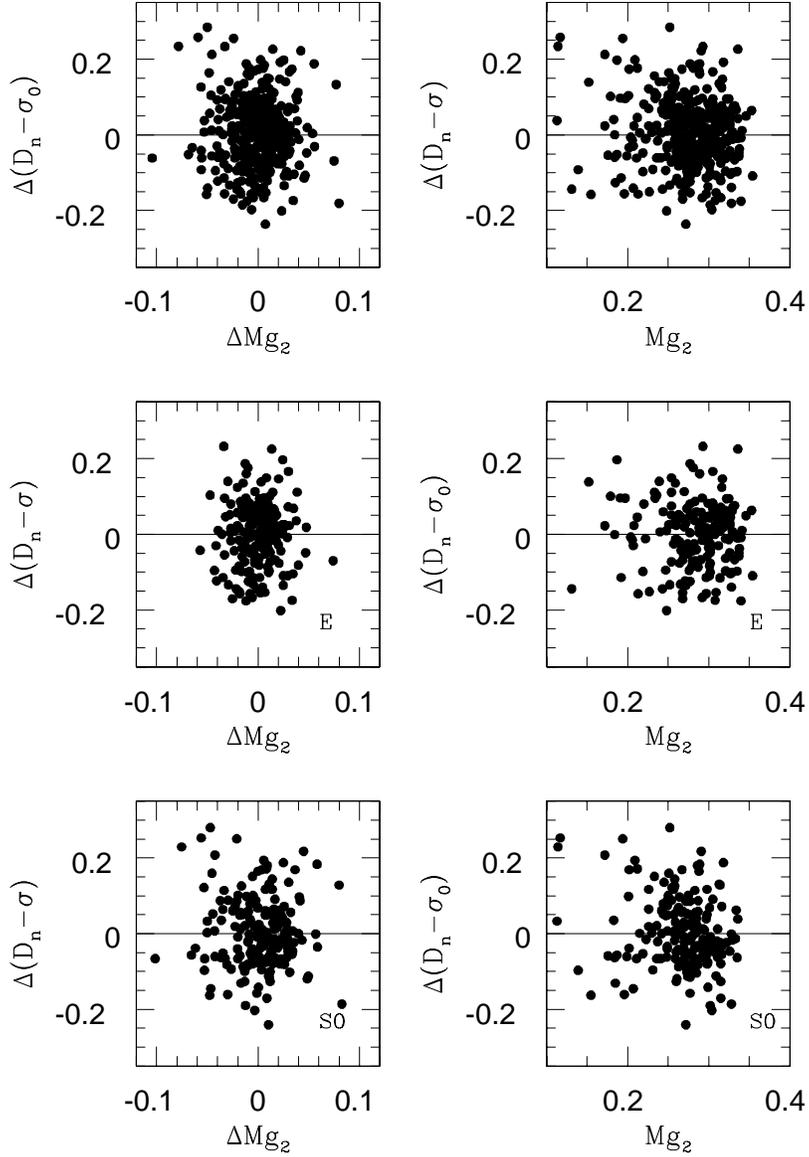,height=15truecm,bbllx=-1truecm,bblly=5truecm,bburx=19truecm,bbury=23truecm }} 
\caption{(Left panels) 
Residuals with respect to the mean 
$D_n-\sigma$ relation versus residuals with respect to the mean 
Mg$_2-\sigma$ relation for the cluster sample as a whole (upper panel),
for the ellipticals (middle panel), and for S0s (lower panel). 
(Right panels) As on the left, but now for the residuals of the 
$D_n-\sigma$ relation versus the measured Mg$_2$ index.}
\label{fig:fitres}
\end{figure}

If differences in stellar populations were important in estimating
distances, one would expect correlations between the \dnsig and
Mg$_2-\sigma$ residuals, $\Delta (D_n-\sigma)$ and $\Delta {\rm
Mg}_2$, respectively, since the latter should reflect either age or
metallicity differences.  The left panels of Figure~\ref{fig:fitres}
show $\Delta (D_n-\sigma)$ versus $\Delta {\rm Mg}_2$ for the cluster
galaxies as a whole (upper panel); for the ellipticals (middle panel);
and for S0s (lower panel), while the right panels show $\Delta
(D_n-\sigma)$ as a function of the Mg$_2$ line index. As can be seen,
there is no obvious correlation between these parameters.  Applying
the Spearman rank test to the data shown in the various panels, we
find that the rank-order correlation coefficients vary from $0.10$ to
$-0.15$, implying significance levels $>0.8$, thereby confirming the
lack of any significant correlation between these quantities.
These results are in agreement with the conclusions of previous
studies (\eg J{\o}rgensen \etal 1996; Colless \etal 1999).

Figure~\ref{fig:DS_group_mg2} shows the data points for each cluster 
and the composite Mg$_2-\sigma$ relation (solid line) given by
the sample as a whole. Open circles indicate S0 galaxies, 
and filled circles, ellipticals. From these panels, it is evident that 
S0 galaxies depart more from the composite relation than ellipticals,
especially at smaller velocity dispersions. Nevertheless, most 
galaxies do lie along the globally derived relation. 
There are some exceptions which were observed by other authors; 
these are listed in Appendix~B.

Our results suggest that differences in stellar populations do not
influence the distance relation enough to mimic peculiar motions.
Furthermore, none of the most discrepant peculiar velocities discussed
in the previous section show evidence that their velocities are caused
by stellar population effects.

\subsection {Environment}

\begin{figure}[t!]
\centering
\mbox{\psfig{figure=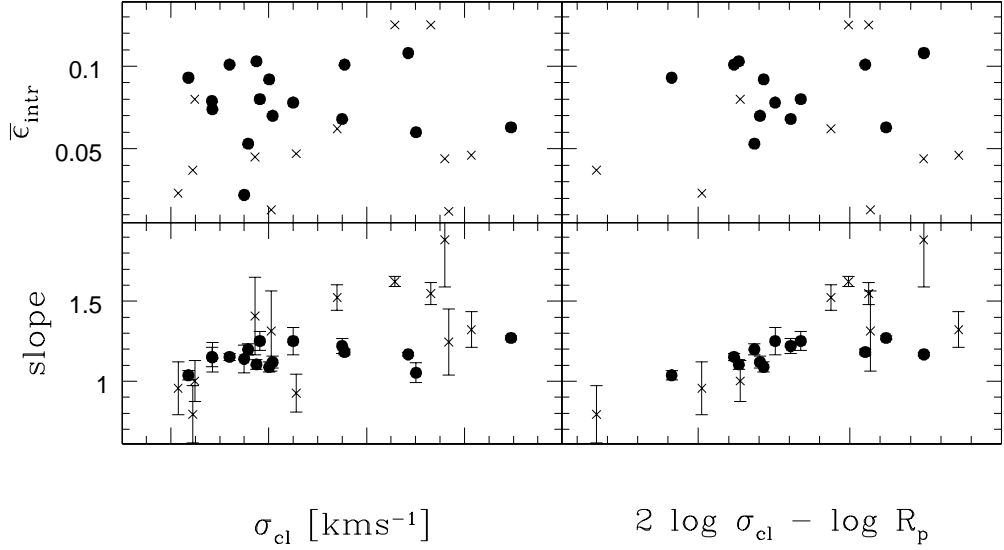,height=9truecm,bbllx=2truecm,bblly=12truecm,bburx=19truecm,bbury=25truecm }}
\caption{
The $rms$ scatter (upper panels) and the slope (bottom panels) of the 
individual cluster $D_n-\sigma$ 
relations versus: (left panels) the measured velocity dispersion of the 
cluster $\sigma_{cl}$; and (right panels) the logarithm of the ratio
$\sigma_{\rm cl}^{2} /R_p$, where $R_p$ is the pair radius defined by
Ramella et al.(1989). Seven clusters taken from the literature, which 
were not identified by the finding algorithm, are not included in the 
right panels of the figure. (Crosses) clusters/groups with a large error 
in the slope ($\gsim 0.1$; see Table~ref{tab:clusindiv}) or which exhibit 
clear evidence of either spatial sub-structure or distinct galaxy populations 
(HMS0122+3305, A2199, Cen30); (filled circles) clusters/groups with reliable 
$D_n-\sigma$ fits.}
\label{fig:fitres4}
\end{figure}

In the literature, there is concern that there may be
cluster-to-cluster environmental differences in the $D_n-\sigma$
method and here we examine this possibility.  Figure~\ref{fig:fitres4}
shows the distribution of the intrinsic scatter
($\bar{\epsilon}_{intr}$) and the slope of the fit as a function of:
a) the measured velocity dispersion of the cluster $\sigma_{\rm cl}$;
and b) the logarithm of the ratio $\sigma_{\rm cl}^{2}/R_p$, where
$R_p$ is the pair radius defined by Ramella, Geller, \& Huchra (1989)
and $\sigma_{\rm cl}^{2}/R_p$ is a rough measure of the projected
cluster surface density.  Clusters/groups with poorly defined slopes
(those with an error in the slope $\gsim 0.1$; see
Table~\ref{tab:clusindiv} and Figure~\ref{fig:DS_group_indiv}) are 
shown as crosses. Also represented by
crosses are the points corresponding to the systems HMS0122+3305,
A2199 and Cen30, where possible membership assignment problems may
affect the determination of the slope (see B01).  Seven clusters from
the literature, for which values for $R_p$ are not available, have not
been included in the right panels of the figure (see B01).

Applying the Spearman rank test the whole sample confirms that there
is no obvious correlation between the intrinsic scatter and the
parameters that characterize the global properties of the clusters.
The rank-order correlation coefficients are $0.01$ and $-0.05$ with
significance levels of $0.96$ and $0.77$, respectively. On the other
hand, a similar analysis of the data shown in the bottom panels of
Figure~\ref{fig:fitres4} might lead one to suspect that the slope of
$D_n-\sigma$ relation depends both on the velocity dispersion and on
the surface density. Taking all the available data points into
consideration the Spearman rank test gives a rank-order correlation
coefficient of $\sim 0.60$ with significance levels of $\sim 5\times
10^{-4}$ in both cases, indicating a strong correlation between the
slope and $\sigma_{cl}$ or $\log \sigma_{\rm cl}^{2}/R_p$. However,
the slope of individual cluster/group relations is in many cases
poorly determined either because of the small number of measured
cluster members or because of interlopers. In fact, if systems with
large errors in the slope (nine systems) and cases where the slope
could be affected by the presence of interlopers (three systems) are
discarded, the Spearman rank-order correlation coefficient decreases to
$\sim 0.40$ corresponding to a significance level of $\sim 0.15$ for
both relations; this shows that the correlation between the slope of
the $D_n-\sigma$ relation and the velocity dispersion or central
surface density is not significant.  Clearly, a more definite test of
this hypothesis requires considerable more data per cluster than
currently available. We should also point out that the Spearman rank
test also shows that there is no obvious correlation between either
the intrinsic scatter or the slope of the cluster's individual
$D_n-\sigma$ relation and the number of observed galaxies. The derived
rank-order correlation coefficients are $0.32$ and $0.11$, yielding
significance levels of $0.4$ and $0.8$, respectively. It should be
emphasized at this point that the good agreement in the peculiar
velocity field obtained from spirals and ellipticals give further
support to the hypothesis of a universal distance-relation.

\begin{figure}[b!]
\centering
\mbox{\psfig{figure=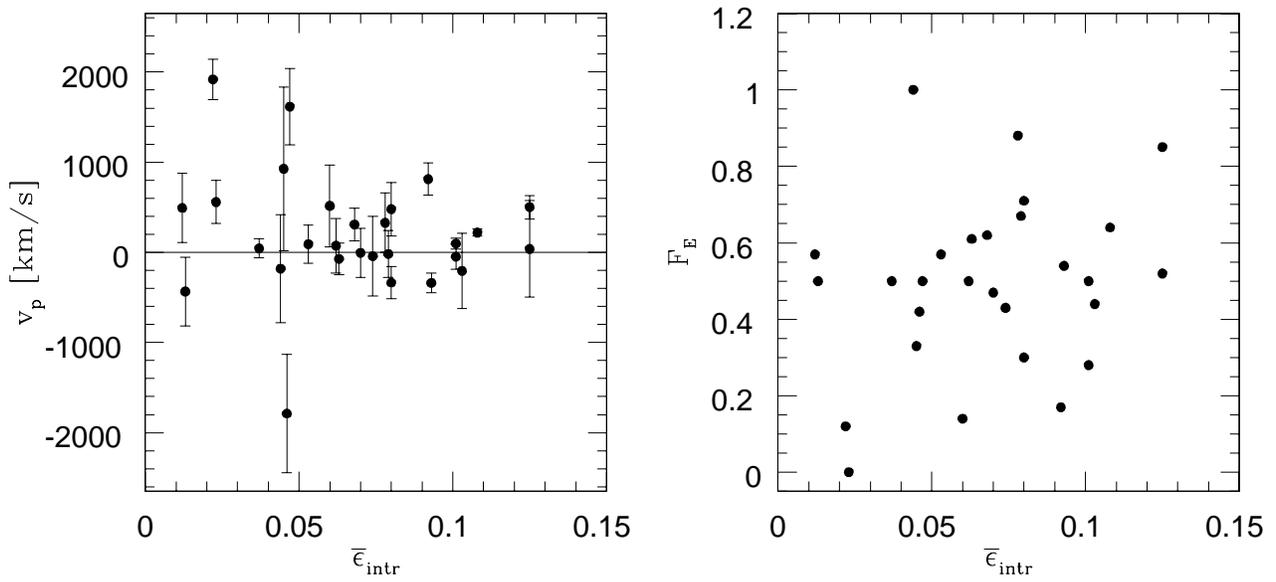,height=9truecm,bbllx=2.5truecm,bblly=15truecm,bburx=19truecm,bbury=25.5truecm }}
\caption{Left panel: peculiar velocities of the 28 clusters as a
function of the amplitude of the scatter of the individual
$D_n-\sigma$ relations of each cluster. Right panel: the fraction of
elliptical to early-type galaxies $F_E = (\rm{N_E}/(\rm{N_{E} +
N_{S0}}))$, for each cluster,
versus the scatter of the individual $D_n-\sigma$ relations.}
\label{fig:velpec_comp2}
\end{figure}

Gibbons \etal (2001) used 20 clusters, of which 15 are in common with
us, to argue that the amplitude of the measured peculiar velocity
correlates with the scatter of the distance relation. The left panel
in Figure~\ref{fig:velpec_comp2} shows the cluster peculiar velocities
of all the clusters in our sample as a function of the amplitude,
$\bar\epsilon_{intr}$, of the intrinsic scatter of the individual
$D_n-\sigma$ relations.  The panel on the right shows the fraction of
elliptical to early-type galaxies ($\rm{N_E}/(\rm{N_{E} + N_{S0}})$)
(right panel), versus $\bar\epsilon_{intr}$.  These plots are similar
to those shown by Gibbons \etal (2001). However, in contrast, we find
no significant correlations in either relation.  The Spearman rank
test gives a rank-order correlation coefficient of $0.21$ with a
significance level of $0.28$ for the $v_p - \bar\epsilon_{intr}$
relation, and correlation coefficient of $-0.16$ with a significance
level of $0.40$ for the relation shown in right panel.

We conclude that cases of poor fits are more likely to be due to
observational limitations rather than reflecting intrinsically
different physical properties.  In summary, we find no compelling
evidence that the peculiar velocities are spurious artifacts.  Rather,
we believe our quoted velocities do measure the motion of clusters
relative to the Hubble flow.

\section{Summary}
\label{summary}

Using new and previously published data for 452 galaxies in 28 
clusters we have derived a bias-corrected \dnsig relation.  
It can be used to measure relative distances of galaxies in the 
recently completed survey of early-type galaxies (da Costa et al. 2000a) 
and to map the peculiar velocity field.  Our main conclusion are:

\begin{enumerate}

\item The slope obtained by combining data for all cluster/groups
does not differ signifcantly from previous determinations.

\item  The scatter is found to be $\sim 0.085$~dex implying a 
distance error of about  20\% per galaxy, comparable to the error 
of FP relations.  Note that $D_n$ is, in general, less sensitive 
to seeing and easier to compute (\eg fits to light profiles are 
not required).

\item Our cluster peculiar velocities are in good agreement with 
other determinations, in particular, with those based on spiral 
TF distances, further supporting the validity of the distance 
indicators.

\item As in previous work we find no evidence for systematic effects
playing a role in the computed peculiar velocities. We believe that
the peculiar velocities we present here are not artifacts but rather a
true measure of the clusters' motions relative to the Hubble flow.

\item Of the 28 clusters in the sample, six show suspiciously large
peculiar velocities (both infall and outflow). Five of these are 
likely due to small-scale dynamical effects, or contamination by
other components. The remaining one is at low galactic latitude 
and may suffer from absorption effects.
Eliminating these clusters we find that the cluster one-dimensional
$rms$ velocity is relative small $239 \pm 46$ kms$^{-1}$, suggesting a
fairly quiescent velocity field, consistent with the estimate 
obtained from the TF data.

\end{enumerate}

The distance relations derived here have been used in previous papers
of this series (da Costa \etal 2000b; Borgani \etal 2000b; Nusser
\etal 2001; Zaroubi \etal 2001) to analyze the peculiar velocity field
traced by early-type galaxies. This sample of early-types, comparable
in size to the SFI sample of field spirals (Haynes \etal 1999a, 1999b), 
allows an independent
analysis of the characteristics of the local velocity field, 
because it uses a different distance relation, and test particles 
which probe a different set of density regimes. The good agreement 
between our early-type cluster sample and the SCI spiral sample 
suggests that it should be possible to merge the ENEAR and SFI 
redshift surveys. This will provide the largest and most homogeneous 
all-sky sample of nearby galaxies available for cosmic flow studies,
and will allow the universality of the results presented here to be
checked directly.

\acknowledgments{The authors would like to thank the referee
for all the helpful comments and all of those who have
contributed directly or indirectly to this long-term project. 
Our special thanks to Ot\'avio Chaves for his many contributions over 
the years. 
We would also like to thank D. Burstein. M. Davis, A. Milone,
M. Ramella, R. Saglia, and B. Santiago for useful discussions and
input.  MB thanks the Sternwarte M\"unchen, the Technische
Universit\"at M\"unchen, ESO Studentship program, and MPA Garching for
their financial support during different phases of this research.  MVA
thanks CNPq for different fellowships at the beginning of the project
and the CfA and ESO's visitor programs for support of visits.  MVA is
partially supported by CONICET and SecyT. LNdC would like to extend
his special thanks to David W. Latham who played a pivotal role at the
early stages of this project. GW is grateful to the Alexander von
Humboldt-Stiftung for making possible a year's stay at the
Ruhr-Universit\"at in Bochum, and to ESO for support for visits to
Garching.  Financial support for this work has been given through
FAPERJ (CNAW, MAGM, PSSP), CNPq grants 201036/90.8, 301364/86-9
(CNAW), 301366/86-1 (MAGM); NSF AST 9529098 (CNAW); ESO Visitor grant
(CNAW). PSP and MAGM thank CLAF for financial support and CNPq
fellowships. Most of the observations carried out at ESO's 1.52m
telescope at La Silla were conducted under the auspices of the
bi-lateral time-sharing agreement between ESO and MCT/Observat\'orio
Nacional.}

\appendix
\section{Notes on the most peculiar cluster velocities}
A2634: one of the most distant clusters in the ENEARc sample 
($cz \sim 9000$ kms$^{-1}$), has a large peculiar velocity 
(it is $>2\sigma$ from the mean defined by our full sample). 
We have no measurements of our own for the this cluster. 
Lucey et al. (1997) re-observed some of its galaxies and concluded 
that the original values of the central velocity dispersion were 
underestimated.  This partially accounts for its large infall velocity. 
Here the peculiar velocity was determined using the 
Lucey et al. (1997) estimates, converted to our system.  
Although our ($\sim -1787$ kms$^{-1}$) is smaller than the number 
reported in Faber et al. (1989) and Lucey et al. (1991) 
($-3400 \pm 600$ kms$^{-1}$), it also disagrees with the more 
recent estimate from Lucey et al. (1997), and with estimate from 
the SCI sample of cluster spirals (G97). 
The Lucey et al. (1997) value is significantly smaller 
($\lsim -700$ kms$^{-1}$) than ours, with the actual value 
depending on the distance relation used (FP or \dnsig).  
In the SCI sample this cluster is nearly at rest relative to the 
Hubble flow.  Hudson et al. (1997), using a sub-sample of the 
Lucey et al. (1997) data set and the FP relation in their paper, 
also find a small infall.  However, using the slope of their 
\dnsig relation, and a zero-point derived from the peculiar 
velocities of the other two clusters we have in common with them, 
A347 and 7S21 (because the zero-point of the $D_n-\sigma$
relation is not reported in that paper), yields a peculiar velocity
of $-1582 \pm 635$ kms$^{-1}$ for A2634, comparable to our value.  
We also note that A2634 has a nearby companion (A2166) at approximately
the same redshift, which may affect membership assignment, and
may explain the large variations in its measured peculiar velocity.

A3381: the most distant cluster/group in the ENEARc sample 
($cz_{cmb} = 11472 \pm 65$ kms$^{-1}$) with only 6 early-type 
galaxies.  This cluster was originally studied by 
J{\o}rgensen et al. (1996) who reported a peculiar velocity of 
$667 \pm 698$ kms$^{-1}$.

AS639: at low galactic latitude ($b \sim 10^\circ$), 
was originally studied by J{\o}rgensen et al. (1996) who found it 
outflowing at $1295 \pm 359$ kms$^{-1}$.  
Correcting to our standard system, we find an amplitude of 
$1615 \pm 433$ kms$^{-1}$.  This may reflect differences in the 
galaxy sample, since we have removed ESO~264G024, 1037-4605,
ESO~264IG030~NED03, and ESO~264IG030~NED02 from it (see Table~8 
in B01).  It may also reflect differences in the adopted distance 
relations.  J{\o}rgensen et al. argued that this large amplitude was
partially due to stellar population differences (section~\ref{stelpop}). 
Using the correlation between the Mg$_2$ line index and the central 
velocity dispersion, they argued that the amplitude of the motion 
was smaller than $\sim 879 \pm 392$ kms$^{-1}$.  
Recently, J{\o}rgensen \& J{\o}nch-S{\o}rensen (1998), 
using additional data, find a peculiar velocity of $838\pm 350$
kms$^{-1}$.  They argue that this is also an overestimate, because 
of evidence for an apparently younger stellar population.
This cluster lies so close to the galactic plane that uncertainties
in absorption correction may be large; these may lead to artificially 
high values of the peculiar velocity.

Cen 30 and Cen 45: their large peculiar velocities can be partially 
explained by the fact that they lie along the same line-of-sight 
and are part of a complex structure. In 
Figure~\ref{fig:DS_group_dist}, Cen30 shows a bi-modal distance 
distribution because it is difficult to assign galaxies to the 
different clumps. While clearly seen in the distance distribution, 
the bimodality is not evident in the redshift distribution in
Figure~\ref{fig:DS_group_cz}. The large positive peculiar velocity 
of Cen45 is likely caused by its infall towards the more massive 
component of the system (\eg Lucey \& Carter 1988). Given the 
complexity of the Centaurus system one should be cautious when
using Cen30 and Cen45.

AS714: suspiciously large amplitude, has 19 members, close to the 
minimum number required to be included in the cluster sample. 
We targeted all 8 early-types in it, of which six are lenticulars.
One of these was excluded from the cluster sample used to
derive the $D_n-\sigma$ relation because it appears to be spiral
(see B01, Table~8). The measured peculiar velocity, 
$559 \pm 245$ kms$^{-1}$, is high. However, the group is located 
in the direction of the GA, which may account for the large amplitude 
(as for Cen30). Because of the complexity of the region, the large 
peculiar velocity can also arise from small-scale dynamical effects, 
such as those in Cen45.

AS753: also in the GA region, shows a large positive peculiar 
velocity of $812 \pm 204$, which is significantly larger than the 
$279 \pm 182$ kms$^{-1}$ obtained by J{\o}rgensen et al. (1996).  
The difference betwen J{\o}rgensen et al. and us is partially due 
to the choices of the galaxy sample, the distance relation and 
our weighting procedures, illustrating how systematic rather than 
random errors can sometimes be responsible for significant 
differences in the measured peculiar velocity of individual clusters.

\section{Notes on the Mg$_2-\sigma$ relation}

\begin{figure}
\centering
\mbox{\psfig{figure=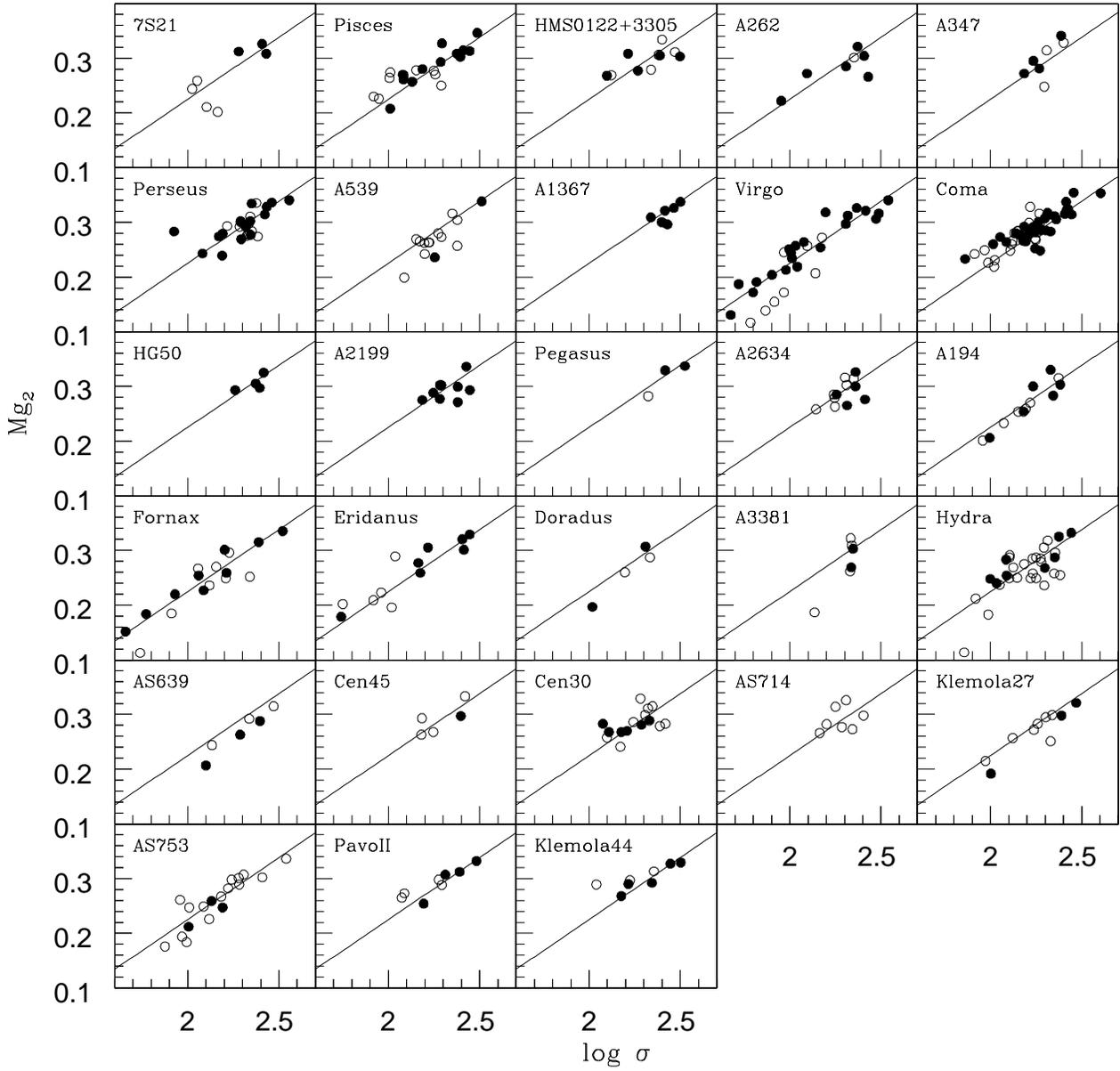,height=15truecm,bbllx=2truecm,bblly=5truecm,bburx=19truecm,bbury=22truecm }} 
\caption{The Mg$_2$ index of each cluster member galaxy versus 
its velocity dispersion. Open circles indicate SO galaxies while
filled circles ellipticals. Solid line shows the derived composite
Mg$_2-\sigma$ relation.}
\label{fig:DS_group_mg2}
\end{figure}

The following galaxies lie off the Mg$_2-\sigma$ relation shown 
in Figure~\ref{fig:DS_group_mg2}.

Perseus: PGC~012423 (Smith et al. 1997) shows a 
higher Mg$_2$ index than that expected from the Mg$_2-\sigma$ relation. 

A539: three galaxies (CGCG~421-015, CGCG~421-017, and 0514+0619a from
J{\o}rgensen et al. 1995b) show a lower Mg$_2$ index. 
They are faint galaxies in a crowded background. 

A3381: PGC~018554 (J{\o}rgensen
et al. 1995b) has a lower Mg$_2$ line index than the expected one. The
spectrum of this galaxy may be affected by the light of a nearby
bright star. 

Hydra: PGC~031765 (the only galaxy observed by us)
also has a low Mg$_2$ index. The spectrum of this galaxy indicates the
presence of weak emission lines and, as pointed out by J{\o}rgensen
et al. (1995a), its image shows the presence of a weak shell.

AS639: all galaxies in this cluster, which is located at low galactic
latitude, seem to lie below the relation (see J{\o}rgensen et al. 1996; 
J{\o}rgensen \& J{\o}nch-S{\o}rensen 1998; see also Section~4). 
All data points for this cluster are from J{\o}rgensen et al. (1995b).

{}


\clearpage

\clearpage

\clearpage

\begin{table}
\begin{center}
\caption {Completeness function coefficients}
\label{tab:param}
\begin{tabular}{lcc}
\\
\hline \hline
Cluster Name & $y_f$ & $\eta$ \\
 (1) &  (2) & (3) \\
\hline
 7S21           &   3.82 &  0.10 \\
\\[-3mm]
 Pisces         &   3.72 &  0.06 \\
\\[-3mm]
 HMS0122+3305   &   4.10 &  0.13 \\
\\[-3mm]
 A262           &   3.95 &  0.15 \\
\\[-3mm]
 A347           &   3.92 &  0.04 \\
\\[-3mm]
 Perseus        &   3.99 &  0.11 \\
\\[-3mm]
 A539           &   3.90 &  0.07 \\
\\[-3mm]
 A1367          &   4.17 &  0.06 \\
\\[-3mm]
 Virgo          &   3.60 &  0.14 \\
\\[-3mm]
 Coma           &   3.80 &  0.09 \\
\\[-3mm]
 HG50           &   3.90 &  0.04 \\
\\[-3mm]
 A2199          &   3.85 &  0.06 \\
\\[-3mm]
 Pegasus        &   4.15 &  0.08 \\
\\[-3mm]
 A2634          &   4.00 &  0.07 \\
\\[-3mm]
 A194           &   3.75 &  0.14 \\
\\[-3mm]
 Fornax         &   3.52 &  0.11 \\
\\[-3mm]
 Eridanus       &   3.45 &  0.05 \\
\\[-3mm]
 Doradus        &   3.78 &  0.06 \\
\\[-3mm]
 A3381          &   4.00 &  0.06 \\
\\[-3mm]
 Hydra          &   3.65 &  0.11 \\
\\[-3mm]
 AS639          &   3.95 &  0.08 \\
\\[-3mm]
 Cen45          &   3.78 &  0.06 \\
\\[-3mm]
 Cen30          &   3.70 &  0.16 \\
\\[-3mm]
 AS714          &   3.90 &  0.04 \\
\\[-3mm]
 Klemola27      &   3.88 &  0.12 \\
\\[-3mm]
 AS753          &   3.68 &  0.09 \\
\\[-3mm]
 PavoII         &   4.00 &  0.08 \\
\\[-3mm]
 Klemola44      &   3.60 &  0.07 \\
\hline
\end{tabular}
\end{center}
\end{table}

\begin{table}
\begin{center}
\caption {Tests of the $D_n-\sigma$ relation}
\label{tab:tests}
\begin{tabular}{lcccc}
\\
\hline \hline
 Objects removed & N$_{\rm remain}$ & $a$ & $b$ & $\bar{\epsilon}$ \\
 (1) & (2) & (3) & (4) & (5)\\
\hline
 A$^{1}$     &  360   &     1.197  &    1.423    &   0.084  \\
\\[-3mm]
 B$^{2}$     &  374   &     1.184  &    1.448    &   0.086  \\
\\[-3mm]
 7S21        &    447  &     1.204    &   1.405    &   0.083 \\  
\\[-3mm]
 Pisces      &    433  &     1.202    &   1.407    &   0.084 \\  
\\[-3mm]
 HMS0122+3305 &   444  &     1.199    &   1.413    &   0.083 \\ 
\\[-3mm]
 A262        &    446  &     1.204    &   1.406    &   0.083 \\ 
\\[-3mm]
 A347        &    447  &     1.203    &   1.407    &   0.083 \\ 
\\[-3mm] 
 Perseus     &    428  &     1.202    &   1.411    &   0.084 \\   
\\[-3mm]
 A539        &    440  &     1.203    &   1.406    &   0.083 \\   
\\[-3mm]
 A1367       &    448  &     1.203    &   1.407    &   0.083 \\   
\\[-3mm]
 Virgo       &    410  &     1.203    &   1.408    &   0.079 \\   
\\[-3mm]
 Coma        &    374  &     1.187    &   1.438    &   0.086 \\   
\\[-3mm]
 HG50        &    447  &     1.203    &   1.405    &   0.083 \\   
\\[-3mm]
 A2199       &    441  &     1.194    &   1.424    &   0.081 \\   
\\[-3mm]
 Pegasus     &    450  &     1.204    &   1.405    &   0.083 \\   
\\[-3mm]
 A2634       &    442  &     1.202    &   1.408    &   0.083 \\   
\\[-3mm]
 A194        &    439  &     1.206    &   1.400    &   0.083 \\   
\\[-3mm]
 Fornax      &    436  &     1.206    &   1.401    &   0.082 \\   
\\[-3mm]
 Eridanus    &    441  &     1.209    &   1.397    &   0.082 \\   
\\[-3mm]
 Doradus     &    450  &     1.202    &   1.408    &   0.083 \\   
\\[-3mm]
 A3381       &    448  &     1.202    &   1.408    &   0.083 \\   
\\[-3mm]
 Hydra       &    415  &     1.204    &   1.404    &   0.081 \\   
\\[-3mm]
 AS639       &    448  &     1.203    &   1.407    &   0.083 \\   
\\[-3mm]
 Cen45       &    446  &     1.203    &   1.408    &   0.083 \\   
\\[-3mm]
 Cen30       &    433  &     1.183    &   1.448    &   0.080 \\   
\\[-3mm]
 AS714       &    447  &     1.203    &   1.407    &   0.083 \\   
\\[-3mm]
 Klemola27   &    444  &     1.203    &   1.408    &   0.083 \\   
\\[-3mm]
 AS753       &    436  &     1.205    &   1.403    &   0.082 \\   
\\[-3mm]
 PavoII      &    442  &     1.201    &   1.412    &   0.083 \\   
\\[-3mm]
 Klemola44   &    436  &     1.204    &   1.406    &   0.082 \\   
\hline
\end{tabular}
\end{center}
\footnotesize
Notes. ---
(1)  peripheral cluster galaxies;
(2) clusters whose individual $D_n-\sigma$ relations differ 
significantly ($\Delta$ slope $\gsim 0.2$) from equation~(\ref{eq:dnsigdir}):
HMS0122+3305, A1367, HG50, A2199, Doradus, A3381, AS639, Cen30, and AS714.
\end{table}

\clearpage

\begin{table}
\begin{center}
\caption {Our determinations of the $D_n-\sigma$ relation}
\label{tab:fits}
\begin{tabular}{lcccc}
\\
\hline \hline
Type & $a$ & $b$ & $\bar{\epsilon}$ & note\\
(1) & (2) & (3) & (4) & (5)\\
\hline
direct & 1.203$\pm0.023$ & 1.406 & 0.085& \\
direct orthogonal fit& 1.414$\pm0.030$ & 0.925 & 0.089& \\
inverse& 1.460$\pm0.028$ & 0.826 & 0.075 & 1\\
\hline
\end{tabular}
\end{center}
\footnotesize
\hspace{2truecm} Notes. ---
(1) the uncertainty in the distances determined using the inverse relation
   is ($a \times \bar{\epsilon}$).
\end{table}

\begin{table}
\begin{center}
\caption {Other determinations of the $D_n-\sigma$ relation}
\label{tab:tablit}
\begin{tabular}{lccccc}
\\
\hline \hline
Source & Type & $a$ & $b$ & $\bar{\epsilon}$ & note\\
 (1) &  (2) & (3) & (4) & (5) & (6)\\
\hline
LC    & direct     & $1.200\pm\ \ \ \ \ $ & -1.679 & 0.090& 1\\
7S    & direct     & $1.200\pm\ \ \ \ \ $ &\ 1.411 & 0.090 & \\
D     & direct     & $1.330\pm\ \ \ \ \ $ & -1.967 & 0.110& 1\\
B96   & direct     & $0.938\pm 0.072$ & - & 0.071& \\
JFK96 & orthogonal & $1.320\pm 0.070$ & - & 0.088&\\
Lc    & direct     & $0.913\pm 0.090$ & -1.019  & 0.075 & 1\\
DCZC97  & direct   & $1.240\pm 0.060$ & -1.080 & 0.080& 2 \\   
HLSS97 & inverse   & $1.419\pm 0.044$ & - & 0.065 & 3\\
GFB    & inverse   & $1.420\pm 0.040$ & - & 0.059 & 3\\ 
\hline
\end{tabular}
\end{center}
\footnotesize
References. ---
LC: Lucey \& Carter (1988);
7S: Lynden-Bell \etal (1988);
D: Dressler \etal (1991);
B96: Baggley (1996);
JFK96 : J\/orgensen \etal (1996);
Lc: Lucey \etal (1997);
DCZC97: D'Onofrio \etal (1997);
HLSS97 : Hudson \etal (1997);
GFB: Gibbons \etal (1998). \\

Notes. --- (1) they used $\log{D_n} = a \log\sigma + b$ with $D_n$ in arcsec. 
Using $D_n = \log{(d_n \times R)}$, where $d_n$ is in 0.1~arcmin, 
one must add $\log R_{{\rm Coma}}-0.778$ to their zero point.\\
(2) as in (1), but substitute $R_{{\rm Coma}}$ for $R_{{\rm Virgo}}$.\\
(3) the uncertainty in the distances determined using the inverse relation
   is ($a \times \bar{\epsilon}$).
\end{table}

\begin{table}
\begin{center}
\caption {Clusters distance and peculiar velocity}
\label{tab:clustervpec}
\begin{tabular}{lrrrrrr}
\\
\hline \hline
\small
Name &n$_{gal}$&l&b & cz$^{CMB}$& R & v$^{CMB}_{pec}$ \\
(1) &  (2) & (3) & (4) & (5) & (6) & (7) \\
\hline
7S21             &   7 & 113.784 & -40.018 &   5500$\pm    90$ &   5542$\pm   441$&   -41$\pm   451$\\
\\[-3mm]
Pisces           &  21 & 127.243 & -30.185 &   4715$\pm    89$ &   4626$\pm   213$&    89$\pm   231$\\ 
\\[-3mm]
HMS0122+3305     &  10 & 130.513 & -28.767 &   4600$\pm   108$ &   4525$\pm   301$&    74$\pm   320$\\ 
\\[-3mm]
A262             &   8 & 136.599 & -25.049 &   4725$\pm    93$ &   4396$\pm   328$&   328$\pm   341$\\ 
\\[-3mm]
A347             &   7 & 141.124 & -17.896 &   5301$\pm   111$ &   4808$\pm   383$&   492$\pm   398$\\ 
\\[-3mm]
Perseus          &  26 & 150.382 & -13.382 &   4799$\pm   133$ &   4490$\pm   185$&   309$\pm   228$\\ 
\\[-3mm]
A539             &  14 & 195.698 & -17.717 &   8636$\pm    75$ &   8119$\pm   457$&   516$\pm   464$\\ 
\\[-3mm]
A1367            &   6 & 234.292 &  73.052 &   6807$\pm    94$ &   6989$\pm   602$&  -181$\pm   609$\\ 
\\[-3mm]
Virgo            &  44 & 283.871 &  74.200 &   1427$\pm    49$ &   1208$\pm    38$&   219$\pm    62$\\ 
\\[-3mm]
Coma             &  80 &  58.301 &  88.285 &   7278$\pm    75$ &   7351$\pm   173$&   -72$\pm   189$\\ 
\\[-3mm]
HG50             &   7 &   0.458 &  49.270 &   1905$\pm    83$ &   2240$\pm   178$&  -334$\pm   197$\\ 
\\[-3mm]
A2199            &  13 &  62.885 &  43.906 &   9108$\pm   111$ &   9069$\pm   530$&    39$\pm   542$\\ 
\\[-3mm]
Pegasus          &   4 &  87.892 & -48.241 &   3202$\pm   108$ &   3635$\pm   383$&  -433$\pm   398$\\ 
\\[-3mm]
A2634            &  12 & 103.402 & -33.161 &   8975$\pm   112$ &  10762$\pm   655$& -1787$\pm   664$\\ 
\\[-3mm]
A194             &  15 & 142.860 & -62.908 &   5074$\pm    60$ &   5079$\pm   276$&    -5$\pm   283$\\ 
\\[-3mm]
Fornax           &  18 & 236.241 & -54.096 &   1330$\pm    36$ &   1234$\pm    61$&    96$\pm    71$\\ 
\\[-3mm]
Eridanus         &  13 & 212.165 & -51.577 &   1488$\pm    28$ &   1827$\pm   106$&  -339$\pm   110$\\ 
\\[-3mm]
Doradus          &   4 & 260.209 & -47.227 &   1073$\pm    36$ &   1028$\pm   108$&    45$\pm   114$\\ 
\\[-3mm]
A3381            &   6 & 240.293 & -22.697 &  11472$\pm    65$ &  10544$\pm   908$&   927$\pm   910$\\ 
\\[-3mm]
Hydra            &  39 & 269.707 &  26.334 &   4055$\pm    95$ &   4103$\pm   138$&   -47$\pm   168$\\ 
\\[-3mm]
AS639            &   6 & 280.534 &  10.908 &   6526$\pm    93$ &   4910$\pm   423$&  1615$\pm   433$\\ 
\\[-3mm]
Cen45            &   8 & 302.553 &  21.659 &   4931$\pm   110$ &   3013$\pm   224$&  1918$\pm   250$\\ 
\\[-3mm]
Cen30            &  21 & 302.023 &  21.852 &   3313$\pm    82$ &   2812$\pm   129$&   500$\pm   153$\\ 
\\[-3mm]
AS714            &   7 & 302.802 &  36.309 &   3576$\pm    49$ &   3017$\pm   240$&   559$\pm   245$\\ 
\\[-3mm]
Klemola27        &  10 & 317.338 &  30.639 &   4881$\pm   102$ &   4402$\pm   293$&   479$\pm   310$\\ 
\\[-3mm]
AS753            &  18 & 319.166 &  26.744 &   4421$\pm    97$ &   3608$\pm   179$&   812$\pm   204$\\ 
\\[-3mm]
PavoII           &  12 & 332.191 & -23.755 &   4266$\pm    63$ &   4285$\pm   261$&   -18$\pm   268$\\ 
\\[-3mm]
Klemola44        &  18 &  25.336 & -75.807 &   8162$\pm    88$ &   8369$\pm   416$&  -206$\pm   425$\\ 
\hline
\end{tabular}
\end{center}
\footnotesize
Notes. --- See Appendix A.
\end{table}

\begin{table}
\begin{center}
\caption {Individual cluster $D_n-\sigma$ relations}
\label{tab:clusindiv}
\begin{tabular}{lccccccc}
\\
\hline \hline
\small				      
Cluster & N$_{\rm gal}$ & $a$ & $\bar{\epsilon}$ &  $\Delta b$ 
& $\bar{\epsilon}_{\Delta}$ & $\bar{\epsilon}_{intr}$ & F$_E$\\
(1) &  (2) & (3) & (4) & (5) & (6) & (7) & (8) \\
\hline
7S21            &   7&  1.150$\pm0.093$ &0.089 &  0.001&0.090&0.074& 0.43\\
\\[-3mm]
Pisces          &  21&  1.200$\pm0.035$ &0.069 & -0.006&0.068&0.053& 0.57\\
\\[-3mm]
HMS0122+3305    &  10&  1.523$\pm0.080$ &0.065 & -0.028&0.074&0.062& 0.50\\
\\[-3mm]
A262            &   8&  1.251$\pm0.085$ &0.089 & -0.013&0.090&0.078& 0.88\\
\\[-3mm]
A347            &   7&  1.246$\pm0.206$ &0.046 & -0.013&0.045&0.012& 0.57\\
\\[-3mm]
Perseus         &  26&  1.220$\pm0.047$ &0.082 & -0.027&0.080&0.068& 0.62\\
\\[-3mm]
A539            &  14&  1.053$\pm0.062$ &0.071 & -0.007&0.072&0.060& 0.14\\
\\[-3mm]
A1367           &   6&  1.883$\pm0.295$ &0.053 & -0.034&0.063&0.044& 1.00\\
\\[-3mm]
Virgo           &  44&  1.168$\pm0.011$ &0.115 &  0.000&0.115&0.108& 0.64\\
\\[-3mm]
Coma            &  80&  1.271$\pm0.012$ &0.071 & -0.002&0.071&0.063& 0.61\\
\\[-3mm]
HG50            &   7&  1.002$\pm0.127$ &0.091 & -0.018&0.093&0.080& 0.71\\
\\[-3mm]
A2199           &  13&  1.549$\pm0.069$ &0.125 &  0.006&0.131&0.125& 0.85\\
\\[-3mm]
Pegasus         &   4&  1.315$\pm0.251$ &0.057 & -0.027&0.058&0.013& 0.50\\
\\[-3mm]
A2634           &  12&  1.323$\pm0.112$ &0.058 & -0.014&0.058&0.046& 0.42\\
\\[-3mm]
A194            &  15&  1.120$\pm0.038$ &0.079 &  0.012&0.080&0.070& 0.47\\
\\[-3mm]
Fornax          &  18&  1.153$\pm0.022$ &0.108 &  0.017&0.108&0.101& 0.50\\
\\[-3mm]
Eridanus        &  13&  1.038$\pm0.029$ &0.093 &  0.035&0.101&0.093& 0.54\\
\\[-3mm]
Doradus         &   4&  0.793$\pm0.179$ &0.042 &  0.001&0.066&0.037& 0.50\\
\\[-3mm]
A3381           &   6&  1.408$\pm0.242$ &0.065 & -0.016&0.067&0.045& 0.33\\
\\[-3mm]
Hydra           &  39&  1.183$\pm0.017$ &0.108 &  0.009&0.108&0.101& 0.28\\
\\[-3mm]
AS639           &   6&  0.927$\pm0.118$ &0.055 & -0.009&0.068&0.047& 0.50\\
\\[-3mm]
Cen45           &   8&  1.140$\pm0.086$ &0.047 &  0.001&0.047&0.022& 0.12\\
\\[-3mm]
Cen30           &  21&  1.623$\pm0.031$ &0.111 & -0.019&0.130&0.125& 0.52\\
\\[-3mm]
AS714           &   7&  0.956$\pm0.166$ &0.048 & -0.001&0.051&0.023& 0.00\\
\\[-3mm]
Klemola27       &  10&  1.252$\pm0.060$ &0.088 & -0.009&0.089&0.080& 0.30\\
\\[-3mm]
AS753           &  18&  1.090$\pm0.031$ &0.097 & -0.006&0.100&0.092& 0.17\\
\\[-3mm]
PavoII          &  12&  1.151$\pm0.061$ &0.086 & -0.006&0.086&0.079& 0.67\\
\\[-3mm]
Klemola44       &  18&  1.105$\pm0.030$ &0.109 &  0.008&0.111&0.103& 0.44\\
\hline
\end{tabular}
\end{center}
\end{table}

\begin{table}
\begin{center}
\caption {Our determination of the Mg$_2-\sigma$ relation}
\label{tab:mg2fit}
\begin{tabular}{lcccc}
\\
\hline \hline
Sample & $N_{\rm gal}$ & $a$ & $b$ & $\bar{\epsilon}$ \\
 (1) & (2) & (3) & (4) & (5) \\
\hline
All   & 369 & 0.226$\pm 0.014$ & -0.227$\pm 0.010$ & 0.021 \\
E     & 186 &                  & -0.226$\pm 0.014$ & 0.019 \\
S0    & 183 &                  & -0.230$\pm 0.015$ & 0.023 \\
\hline
\end{tabular}
\end{center}
\end{table}

\end{document}